\documentclass[11pt]{article}
\usepackage{epstopdf}                                                          
\usepackage[a4paper,hmarginratio=1:1,vmarginratio=2:3,totalwidth=15.6cm,
  totalheight=22.5cm]{geometry}
\usepackage{bm,epstopdf,epsfig,amsmath,amssymb,
amsfonts,colordvi,wrapfig,comment,cancel,verbatim,slashed}
\usepackage{graphicx,graphics}
\usepackage[font=md,captionskip=8pt]{subfig}
\usepackage[usenames,dvipsnames]{color}
\usepackage[noadjust]{cite}
\usepackage{xcolor} 
\usepackage[utf8]{inputenc}
\usepackage{setspace}

\usepackage[utf8]{inputenc}

\newcommand{\dg}{{g^*}}
\newcommand{\dgp}{g^{* \prime}}
\newcommand{\dphi}{{\phi^*}}
\newcommand{\dphip}{\phi^{* \prime}}

\newcommand{\w}{\omega}

\allowdisplaybreaks

\newcommand{\cA}{{\cal A}}

\newcommand{\cD}{{\cal D}}
\newcommand{\cE}{{\cal E}}
\newcommand{\cF}{{\cal F}}

\newcommand{\cJ}{{\cal J}}

\newcommand{\cO}{{\cal O}}

\newcommand{\be}{\begin{equation}}
\newcommand{\ee}{\end{equation}}
\newcommand{\bea}{\begin{eqnarray}}
\newcommand{\eea}{\end{eqnarray}}
                
\newcommand{\ra}{\rightarrow}  
\newcommand{\Ra}{\Rightarrow}

\newcommand{\baa}{\begin{array}}
\newcommand{\eaa}{\end{array}}

\long\def\symbolfootnote[#1]#2{\begingroup
\def\thefootnote{\fnsymbol{footnote}}\footnote[#1]{#2}\endgroup}

\usepackage[subfigure]{tocloft}

\begin{document} 
\begin{flushright}
\end{flushright}

\thispagestyle{empty}
\vspace{2.cm}
\begin{center}
\vspace{1.5cm}

{\Large\bf Weyl conformal geometry  vs Riemannian geometry }

\bigskip

{\Large \bf  of Weyl  gauge invariant (dressed) metric}

 \vspace{1.5cm}
 
 {\bf D. M. Ghilencea}
 \symbolfootnote[1]{E-mail: dumitru.ghilencea@cern.ch}
 and
 {\bf V.-M.  Mandric}
 \symbolfootnote[2]{E-mail: vlad.mandric@theory.nipne.ro}
 
\bigskip 

{\small Department of Theoretical Physics, National Institute of Physics
 \smallskip 

 and  Nuclear Engineering (IFIN), Bucharest, 077125 Romania}
\end{center}

\begin{abstract}\noindent
Weyl conformal geometry is the natural underlying geometry of 
gauge theories of  Weyl group (of dilatations and Poincar\'e symmetry),
such as Weyl quadratic gravity and its  generalisation, 
Weyl-Dirac-Born-Infeld  action (WDBI). These are local, Weyl-anomaly free
(quantum) gauge theories of gravity.
We describe Weyl gauge symmetry from a more familiar Riemannian
view of Weyl gauge invariant  dressed fields by the Wilson line of
dilatations. Weyl geometry can then be seen as Riemannian geometry
of  non-local dressed metric ($g_{\mu\nu}^*$), at the ``cost''
of (gauge-induced) non-commutativity in the UV, due to Wilson line. 
Then Weyl quadratic gravity and WDBI actions of Weyl geometry, which  are 
Weyl gauge invariant in $d$ dimensions, have the same expression  in
Riemannian geometry defined by $g^*_{\mu\nu}$. This is a {\it non-local}
map and dual description of the two geometries and actions in the
symmetric phase. Unlike for the metric, the equation of motion
of Weyl gauge field ($\omega_\mu$) does not commute with the dressing
of the metric. Quantum non-locality (and possibly entanglement) and
non-commutativity are  (gauge-invariant) {\it physical}
 artefacts of ``translating'' (local)
 Weyl geometry and Weyl gauge covariance into our real-world
 Riemannian geometry of Weyl gauge invariant observables,
 and they are  evidence of Weyl gauge symmetry.
 At lower energies,  $\omega_\mu$ becomes massive, can decouple and
Einstein-Hilbert action and commutativity are recovered.
The case of a light $\omega_\mu$ is also discussed.
\end{abstract}

\newpage

{ \tableofcontents}

\section{Motivation}

In 1918 Weyl introduced a ``true local geometry'' now known as
Weyl conformal geometry \cite{Weyl1,Weyl2,Weyl3} that  is a natural
generalisation of Riemannian geometry.
The associated action, quadratic in curvatures, was the first ever
gauge theory constructed,
of the Weyl group of dilatations and Poincar\'e symmetry.
This action is  a  vector-tensor gauge theory of gravity,
to which we refer as Weyl (quadratic) gauge theory of gravity\footnote{Weyl's
original physical interpretation of this theory as a theory of gravity
and electricity was short-lived, since the Weyl gauge boson of
dilatations $\omega_\mu$ is not the real photon as Weyl had initially thought,
but a vector field that together with the metric mediates gravitational interactions.}.
Its underlying Weyl geometry  can be seen as a Riemannian geometry\footnote{
By an abuse of language, we actually mean pseudo-Riemannian/Weyl
geometries throughout the paper.} ``covariantised'' with respect to gauged
dilatations. For  brief reviews of recent progress in Weyl gauge theory of gravity
see \cite{Ghilencea:2026,review} and for a historical review see \cite{Scholz}.

Our interest in this geometry is motivated by the physical interpretation of
Weyl conformal geometry as a candidate for a (quantum) gauge theory of gravity
\cite{review}, as outlined here.
First, in its Weyl gauge covariant formulation, Weyl conformal
geometry is {\it metric}, see \cite{Dirac} and more recent
\cite{DG1,Lasenby,non-metricity,CDA2,CDA,Ghilen0},
and thus it is  physically relevant, contrary to a century-old view due to  Einstein
\cite{Weyl1}\footnote{A century ago, Einstein's argument  \cite{Weyl1} against
the physical relevance of
Weyl  geometry invoked its non-metricity ($\tilde\nabla_\mu g_{\alpha\beta}\not=0$),
but at that time it was not known:  (1) the role of (Weyl) gauge covariance
 (understood later, in  modern gauge theories)
which - when respected - renders this geometry metric
\cite{Dirac,DG1,Lasenby,non-metricity,CDA2,CDA}
and: (2) that the Weyl gauge field $\omega_\mu$ of dilatations
is actually a massive gauge field with mass $\alpha M_p$ ($M_p$ Planck mass), so it
decouples; non-metricity effects
due to it are then suppressed by its large $m_\omega\sim M_p$  if $\alpha\sim \cO(1)$
\cite{Ghilen0} or by an  ultraweak Weyl gauge coupling  $\alpha\ll 1$;
current non-metricity
lower bound is $m_\omega\sim 10^{-14}$ eV \cite{saridakis}.}.
The associated quadratic action has a Stueckelberg breaking
of Weyl gauge symmetry in which the Weyl gauge boson of dilatations $\omega_\mu$ becomes
massive \cite{Ghilen0} and decouples; then Weyl connection and  geometry
become Riemannian and  below the mass of $\omega_\mu$, $m_\omega\!\sim \! M_p$ ($M_p$:
Planck scale), one recovers  Einstein-Hilbert action and a
positive cosmological constant $\Lambda\!>\!0$.

Weyl gauge theory is the only  gauge theory of a spacetime symmetry
(beyond Poincar\'e symmetry) that has a physical (dynamical) gauge
boson\footnote{Gauging the larger, full conformal
group instead of  Weyl group does not lead to a {\it true} gauge theory since
its action cannot have dynamical (physical) gauge bosons of
special conformal symmetry or gauged dilatation~\cite{KNT}.},
is Weyl-anomaly free \cite{DG1,review}
and has  all mass scales of geometric origin
\cite{SMW,non-metricity,Ghilen0}. This means the theory has an exact geometric
interpretation.  SM admits a natural embedding in 
Weyl  geometry with no new degrees of freedom beyond those of Weyl geometry and
SM \cite{SMW}. There is 
successful Starobinski-Higgs inflation \cite{WI3,WI1,Tang} and a good
fit of SPARC galactic rotation curves  \cite{Harko,Harko1}.
A more fundamental theory, known as Weyl-Dirac-Born-Infeld (WDBI)
gauge theory of Weyl group exists in arbitrary $d$ dimensions
in Weyl geometry, with the unique feature that it
does {\it not} need a UV regulator \cite{DBI,WDBI}.
In $d=4-2\epsilon$, the leading order (in $\xi^2\!\sim\!\Lambda/M_p^2$)
of this WDBI action recovers exactly the  Weyl gauge invariant {\it geometrically
regularised} Weyl quadratic gravity.

Motivated by this interesting physics, the goal of this paper is to show
a new perspective on  Weyl conformal geometry and its  action
in the Weyl gauge symmetric phase, from  the more familiar Riemannian geometry
in which we actually  live. Normally, Riemannian geometry is used to describe the broken
phase of Weyl geometry,
in which Einstein-Hilbert gravity is recovered \cite{Ghilen0}.
The question is whether we can also describe  the symmetric phase in
a Riemannian-like picture,  using   {\it gauge invariant} fields.
This can give a map connecting the two  geometries.
This is indeed possible using so-called dressed fields by Wilson lines
\cite{Dirac2,Francois,mandelstam,mandelstam2,terning,Francois2,Wilson}
which are  gauge invariant (rather than covariant) fields.
We use a  dressed metric ($g_{\mu\nu}^*$), dressed dilaton ($\phi^*$) and dressed
curvatures; these are fields dressed by the Wilson line integral
(for a review \cite{faraji}) of the Weyl gauge field of dilatations ($\omega_\mu$).

In gauge theories of gravity such as those with Weyl gauge symmetry,
we need solutions (for the metric, etc) that respect this symmetry,
so that conclusions are physical.
The use of dressed fields is then   an  advantage compared 
to the  Weyl gauge covariant solutions of Weyl geometry or to their
counterparts in traditional Riemannian picture  in which ordinary curvature
fields  transform non-covariantly, in a complicated way.
Being gauge invariant, the dressed  fields are 
physically relevant and their use  avoids
gauge-dependent artefacts and results (e.g. conclusions regarding
the presence or absence of metric singularities, etc).

In this approach we  show Weyl geometry with its Weyl gauge symmetry
may be seen as Riemannian geometry  of  (dressed) metric $g_{\mu\nu}^*$
at the cost of ultraviolet (UV)
non-commutativity ($\propto\! F_{\mu\nu}$) of partial derivatives
when acting on dressed fields.
The Wilson line essentially  moves the dependence on $\omega_\mu$
of Weyl connection into a path-dependent dressed metric and so
Weyl connection/geometry become Riemannian. The dressing
maps the action of Weyl geometry in $d$ dimensions,
to Weyl gauge invariant  action {\it of same expression} in Riemannian geometry
of $g^*_{\mu\nu}$, but with  UV non-commutativity.
The Wilson line sums quantum fluctuations of $\omega_\mu$ (background
$\overline\omega_\mu\!=\! 0$), so dressing  and non-commutativity
are  quantum effects. The map applies to both Weyl quadratic gravity  and
 WDBI action.  We thus have a {\it non-local}  map  and dual description of
 the two geometries and  actions in\,the\,symmetric phase. 
 We study the impact of geometry dressing on equations of motion: dressing
 the metric commutes with the equation of motion of the metric, but
 not with the equation of motion of $\omega_\mu$. We then detail the
 physical implications of our results. Having understood
 quantum non-locality as consequence of Weyl gauge invariance,
 we argue quantum entanglement may be a particular case of it, for a light $\omega_\mu$.
 In\,the\,broken phase massive $\omega_\mu$
decouples and commutativity is restored.

  \section{Brief review of Weyl geometry and its action}
  \label{2}

In this section we review Weyl conformal geometry  \cite{Weyl1,Weyl2,Weyl3}
and its associated Weyl gauge theory in the Weyl covariant (metric) formulation,
following \cite{DG1,review}.  Weyl conformal geometry (WG)
can be regarded as a gauge theory of the Weyl group of dilatations and Poincar\'e
symmetry \cite{Tait,CDA,AC}.  We work in $d$ dimensions; for physics, we
restrict the analysis to $d=4-2\epsilon$ ($\epsilon\ll 1$)
or $d=4$. Weyl geometry is defined by classes of equivalence of
the metric $g_{\mu\nu}$ and Weyl gauge field $\omega_\mu$ (or,
equivalently,  Weyl connection), related by Weyl gauge transformations:
\bea\label{WGS}
 &\quad&
 g_{\mu\nu}^\prime=\Sigma^2 
 \,g_{\mu\nu},\qquad
 \w_\mu'=\w_\mu - \partial_\mu\ln\Sigma, 
\qquad
\sqrt{g'}=\Sigma^{d} \sqrt{g}.
\eea

\medskip\noindent
where we set to 2 the charge of the metric.
Here $\ln\Sigma(x)$ ($\Sigma(x)>0$) is the gauge
function; in the limit $x\ra x_0= \pm \infty$, we expect
$\omega_\mu\ra 0$ and then $\Sigma\ra$ constant
(we set $\Sigma(x_0)=1$).
The definition is completed by  so-called ``non-metricity'' condition:
\bea\label{conn}
\tilde\nabla_\mu g_{\alpha\beta}= - 2 \,\omega_\mu g_{\alpha\beta},
\quad
\textrm{where}
\quad
\tilde \nabla_\lambda g_{\mu\nu}=\partial_\lambda g_{\mu\nu} -
\tilde\Gamma^\rho_{\lambda\mu} g_{\rho\nu}-\tilde\Gamma^\rho_{\lambda\nu}g_{\rho\mu}.
\eea
Assuming a symmetric connection,
$\tilde\Gamma_{\mu\nu}^\rho=\tilde\Gamma_{\nu\mu}^\rho$, then  from (\ref{conn})
\medskip
\bea\label{tGamma}
\tilde\Gamma_{\mu\nu}^\rho(g,\omega)=\Gamma_{\mu\nu}^\rho(g)
+\big[\delta_\mu^\rho \,\omega_\nu + \delta_\nu^\rho\,\omega_\mu -\omega^\rho\,g_{\mu\nu}\big],
\eea
with Levi-Civita connection  $\Gamma(g)$,
\bea\label{LC}
\Gamma_{\mu \nu}^\rho(g) = (1/2) g^{\rho \lambda}(\partial_\mu g_{\nu \lambda}
+ \partial_\nu g_{\mu \lambda} - \partial_\lambda g_{\mu \nu}) \, .
\eea

\medskip\noindent
 $\tilde\Gamma$ is Weyl gauge  invariant
(the variations of $g_{\mu\nu}$ and $\omega_\mu$ compensate each other).
 If $\omega_\mu \ra 0$ (e.g. if $\omega_\mu$ becomes massive and decouples)
 then $\tilde\Gamma\ra \Gamma$ and Weyl geometry becomes Riemannian.

\medskip
\bigskip\noindent
{\bf $\bullet$ Non-metric affine formulation ($\tilde\nabla$)}

\bigskip\noindent
In this historical formulation of WG one is using $\tilde\nabla_\mu (\tilde\Gamma)$
to define the Riemann curvature tensor of Weyl geometry,
$\tilde R^\rho{}_{\sigma \mu \nu}$, via the commutator of
 $\tilde\nabla_\mu$ acting on a vector $T^\rho$
\medskip
\begin{equation}\label{t1}
[\tilde \nabla_\mu, \tilde \nabla_\nu]\, T^\rho = \tilde R^\rho{}_{\sigma \mu \nu}\, T^\sigma,
\end{equation}
which means
\begin{equation}
\tilde R^\rho{}_\sigma{}_{\mu \nu} = \partial_\mu \tilde \Gamma^\rho_{\nu \sigma}
  - \partial_\nu \tilde \Gamma^\rho_{\mu \sigma} + \tilde \Gamma^\rho_{\mu \tau}
  \tilde \Gamma^\tau_{\nu \sigma} - \tilde \Gamma^\rho_{\nu \tau} \tilde \Gamma^\tau_{\mu \sigma}\, .
\label{curvature-1}
\end{equation}

\medskip\noindent
This expression is similar to Riemann tensor in Riemannian geometry,
in terms of  $\Gamma$.
With (\ref{tGamma}) one computes the Ricci curvature tensor  of WG (Weyl-Ricci)
$\tilde R_{\sigma\nu}\equiv \tilde R^\rho_{\,\,\,\sigma\rho\nu}$
 and  Weyl-Ricci scalar: $\tilde R\equiv g^{\sigma\nu} \,\tilde R_{\sigma\nu}$.
 Since $\tilde\Gamma$ is invariant under (\ref{WGS}),
  $\tilde R^\rho_{\,\,\,\sigma\mu\nu}$
 and $\tilde R_{\sigma\nu}$ are  also Weyl gauge invariant;
 $\tilde R$ is Weyl gauge covariant (charge $-2$) due to 
 $g^{\mu\nu}$ in its definition, so  $\tilde R'=\Sigma^{-2} \tilde R$.
 Their expressions in terms of their
 Riemannian geometry counterparts are shown in (\ref{ss1}).

This is the affine non-metric formulation, commonly used
in the literature. This formulation has the problem that it
is not Weyl gauge covariant: while Weyl-Riemann and Weyl-Ricci tensors are
Weyl gauge invariant and $\tilde R$ is Weyl gauge covariant,
their $\tilde\nabla_\mu$ derivatives are not gauge covariant. For example, 
$\tilde \nabla'_\mu \tilde R'\not= \Sigma^{-2}\tilde\nabla_\mu\tilde R$.
The lack of Weyl gauge covariance of this formulation means that
its results are  not physical, except  when no terms
with $\tilde\nabla_\mu$  are present. Moreover,  to perform calculations
one must go to the (metric) Riemannian picture.
Fortunately, there exists an alternative formulation of Weyl geometry,
discussed next.

\vspace{0.5cm}
\noindent
{\bf $\bullet$  Weyl gauge covariant, metric  formulation ($\hat\nabla$)}

\bigskip\noindent
In Weyl geometry regarded as a physical gauge theory of Weyl group,
Weyl gauge covariance must be respected. 
There  does exist  a  Weyl gauge covariant formulation ($\hat\nabla_\mu$)
which is automatically  {\it metric}, see
\cite{Dirac,DG1,non-metricity, Lasenby,CDA,CDA2,Ghilen0}.
This is inspired by noticing that in  (\ref{conn})
one can redefine $\tilde\nabla_\mu$
to include the right-hand side ($\propto\omega_\mu$) into a new,
Weyl gauge covariant derivative ($\hat\nabla_\mu$); this is possible since Weyl
charge of the metric is 2 and then the
theory becomes metric with respect to $\hat\nabla_\mu$.
To detail, 
for a curvature tensor or scalar  $T$ of space-time charge
$\tilde q_T$, that transforms under (\ref{WGS}) as $T'=\Sigma^{\tilde q_T}\,T$,
one defines a Weyl gauge covariant derivative $\hat\nabla_\mu$ \cite{DG1}
\smallskip
\bea\label{qq}
\hat \nabla_\mu T
\equiv (\tilde\nabla_\mu +\tilde q_T\, \w_\mu)\, T\qquad
\Ra\qquad \hat\nabla_\mu' T'=\Sigma^{\tilde q_T}\, \hat\nabla_\mu T,
\eea

\smallskip\noindent
and $\hat\nabla_\mu g_{\alpha\beta}=0$, so the theory is also metric.
Since $\hat\nabla_\mu$ depends on the charge $\tilde q_T$,
no $\hat\Gamma$  can be defined for all fields $T$  on which it acts, hence
this formulation is  not affine.

The Weyl-Riemann tensor $\hat R^\lambda_{\,\,\,\mu\nu\sigma}$ defined 
using new $\hat\nabla_\mu$ (instead of old $\tilde\nabla_\mu$) is \cite{CDA2,CDA,AC}
\bea\label{curvature-hat}
    [\hat\nabla_\mu, \hat\nabla_\nu]\,v^\rho=
    \tilde  R^\lambda_{\,\,\,\sigma\mu\nu}\,v^\sigma+\tilde q_v \,F_{\mu\nu} v^\rho
    =\hat R^\lambda_{\,\,\sigma\mu\nu}\,v^\sigma+q_v\,\hat F_{\mu\nu} \,v^\rho,
    \eea

\medskip\noindent
where $F_{\mu\nu}=\partial_\mu\omega_\nu-\partial_\nu\omega_\mu=\hat F_{\mu\nu}$.
Here  $v^\mu=e^\mu_a\,v^a$ is a vector
with space-time Weyl charge $\tilde q_v$ and tangent space charge $q_{v}$;
 $\tilde q_v=q_v-1$ since $e^\mu_a$ has charge -1.
With this, one finds  the  Weyl-Riemann tensor $\hat R^\mu_{\,\,\,\nu\rho\sigma}$,
Weyl-Ricci tensor $\hat R_{\mu\sigma}=\hat R^{\lambda}_{\,\,\,\mu\lambda\sigma}$ and
Weyl-Ricci scalar $\hat R=\hat R_{\mu\sigma} g^{\mu\sigma}$.
The relation of this Weyl covariant (metric) formulation
  to the previous non-metric one is:
  \medskip
  \bea\label{relations-curvatures}
  \hat R^\rho_{\,\,\sigma\mu\nu}=\tilde R^\rho_{\,\,\sigma\mu\nu}
  -\delta_\sigma^\rho\,\hat F_{\mu\nu},
  \quad \hat R_{\sigma\nu}=\tilde R_{\sigma\nu} -\hat F_{\sigma\nu},
  \quad
  \hat R=\tilde R.
  \eea
  
  \medskip  \noindent
The relations of these curvatures to their Riemannian geometry
 counterparts are found in  Appendix~\ref{A}.  
  One also shows \cite{DG1} (eq.A-25) that in this Weyl gauge covariant formulation,
 the Weyl tensor $\hat C^\mu_{\,\,\nu\rho\sigma}$ associated to
 $\hat R^\mu_{\,\,\,\nu\rho\sigma}$ is identical to its Riemannian geometry version
 ($C^\mu_{\,\,\nu\rho\sigma}$) i.e.  $\hat C^\mu_{\,\,\nu\rho\sigma}
 =C^\mu_{\,\,\nu\rho\sigma}$.
Finally, under (\ref{WGS}), we have \cite{DG1}
\medskip
\bea
\label{WGS3}
&& \hat R^\prime=\Sigma^{-2} \hat R,\qquad
\hat R^{\prime}_{\mu\nu}=\hat R_{\mu\nu},\qquad
    \hat  R^{\prime\,\sigma}_{\,\,\,\mu\nu\rho}=\hat R^\sigma_{\,\,\,\mu\nu\rho},\qquad
\nonumber\\[5pt]
&&\hat\nabla_\mu' \hat R^\prime=\Sigma^{-2}\hat \nabla_\mu \hat R,\qquad
\hat\nabla^\prime_\mu\hat\nabla^{\prime}_\nu \hat R'
=\Sigma^{-2} \hat \nabla_\mu\hat\nabla_\nu \hat R, \quad 
\hat\nabla_\alpha' \hat R_{\mu\nu}'=\hat\nabla_\alpha \hat R_{\mu\nu},\,\,\,
\text{etc.}
\nonumber\\[5pt]
&& X^\prime=\Sigma^{-4} X, \qquad
X  =\hat R_{\mu\nu\rho\sigma}^2, \, \,\hat R_{\mu\nu}^2,\,\,\hat R^2,
 \,\, \hat C_{\mu\nu\rho\sigma}^2,\,\, \hat G,
\,\, \hat F_{\mu\nu}^2,
\eea

\medskip\noindent
where $\hat G$ is the  Euler-Gauss-Bonnet term (see Appendix) which in $d=4$
is a total derivative.

This is the only formulation that is Weyl-gauge-covariant and
metric. In this formulation we can say that Weyl conformal
geometry is simply Riemannian geometry covariantised with respect to
dilatations symmetry \cite{DG1,review}
(there is a third  equivalent formulation, metric, non-covariant, with
torsion \cite{CDA}, obtained from the covariant one by a projective
transformation).

\vspace{0.5cm}
\noindent
{\bf $\bullet$ Weyl gauge theory action}

\bigskip\noindent
The Weyl gauge covariant formulation is important
since it means it is {\it  physical} and   can be
used in applications, while its metricity means one does not have
to rely on the (metric) Riemannian picture to do calculations.
The  action is {\it  the same} in all these formulations, up to
a re-definition of coupling $\alpha$ (all couplings $\xi, \eta, \alpha<1$ are
dimensionless) \cite{Weyl2,Ghilen0,SMW,review}
\bea\label{WWW}
S_{\bf w}=\int d^4 x \,\sqrt{g}\,
\Big\{\frac{1}{4!\,\xi^2}\hat R^2
-\frac{1}{\eta^2} \,\hat C_{\mu\nu\rho\sigma}^2
-\frac{1}{4\,\alpha^2}\, \hat F_{\mu\nu}^2+\hat G
\Big\}
\eea

This action has  interesting properties. To analyse it, it helps to linearise
the $\hat R^2$ term by replacing $\hat R^2\ra -2\phi^2\hat R-\phi^4$.
The new action so obtained gives an equation of motion
$\phi^2=-\hat R$, which replaced back in the new action gives (\ref{WWW}),
so the two actions are classically equivalent. The field $\ln\phi$ transforms with a shift
under (\ref{WGS}) and plays the role of would-be-Goldstone of Weyl gauge symmetry
(``dilaton'' ghost). Further,  $\omega_\mu$
becomes massive (\`a la Stueckelberg) by eating this $\ln\phi$ field
propagated by $\hat R^2$ \cite{Ghilen0}. So the term  $\hat R^2$
drives the spontaneous breaking of Weyl gauge symmetry that is eating the ghost;
one  obtains the Einstein-Hilbert action plus Proca action of $\omega_\mu$
and a  cosmological constant $\Lambda>0$, with mass scales
\cite{Ghilen0,SMW,review}
\bea\label{scales}
\Lambda\equiv \frac14\,\langle\phi\rangle^2,\qquad
M_p^2\equiv \frac{\langle\phi^2\rangle}{6\,\xi^2},\qquad
m_\omega^2\equiv 6 \alpha^2\,M_p^2.
\eea
$M_p$ is the Planck scale, $\Lambda$ is
cosmological constant and $m_\omega$ is the mass of $\omega_\mu$,
all fixed by  $\xi$, $\alpha$ and $\langle\phi\rangle$.
All masses have then a geometric origin.
We have $\xi\sim \Lambda/M_P^2\ll 1$.
Regarding coupling $\alpha$ which is a parameter of the theory,
one can set $\alpha\sim \cO(1)$ then $m_\omega\sim M_p$ and $\omega_\mu$
decouples and Weyl geometry
($\hat\nabla_{\mu}(\tilde\Gamma)$) becomes Riemannian ($\nabla_\mu(\Gamma)$),
see (\ref{tGamma}), (\ref{LC}).
One is left below $M_p$, in the broken phase,
with Einstein-Hilbert action  \cite{Ghilen0}\footnote{
For suitable boundary conditions \cite{Maldacena,Hell}
the  term $\hat C_{\mu\nu\rho\sigma}^2$ in $S_{\bf w}$
does not generate a spin-2 ghost state in the spectrum,
which is a very important result (otherwise,
without such conditions, in the presence of Einstein-Hilbert  term
in the action, this state  has  a mass near Planck scale $\eta M_p\!\sim\!  M_p$ 
for a  natural $\eta\!\sim\! O(1)$, and is integrated out;
 there is unitarity violation (very small), but
one does not produce a negative norm state
\cite{Hawking}).}.

In principle,  $\omega_\mu$ can also  be {\it very light} since it has no 
couplings to SM except to the Higgs \cite{SMW} to which the coupling is
$\alpha\sim m_\omega/M_p$. 
Currently, values of $m_\omega\sim 10^{-13}$ eV are allowed
\cite{oscar,caputo,saridakis,ER}, giving 
$\alpha\sim 10^{-40}$ for which phenomenological constraints are
avoided\footnote{A gauge
kinetic mixing of $\omega_\mu$ with SM hypercharge gauge field \cite{SMW} is 
possible, but such coupling violates $C$-parity in the gauge sector
($\omega_\mu$ is C-even); this mixing coupling is  unnatural in the gauge sector of SM;
it must also be ultraweak to avoid $Z$-boson mass corrections \cite{SMW}.}.

Briefly,  action $S_{\bf w}$  is an ultraviolet (UV) completion of
Einstein-Hilbert  gravity into a  gauge theory of the Weyl group.
There is a conserved scale current $\hat\nabla_\mu j^\mu=0$,
$j_\mu\sim \hat\nabla_\mu \hat R$ which generalises
a similar current in globally scale invariant case \cite{F1,F2,F3,F4,Bellido}.
Further,  SM with a vanishing Higgs mass
parameter, being scale invariant, can  naturally be embedded in
this geometry, with no new degrees of freedom beyond those of SM
and WG \cite{SMW}. Successful Starobinski-Higgs inflation is possible
\cite{WI3,WI1,Tang}, which is a gauged version of Starobinsky inflation.
These results motivated our analysis here  of other physical implications of WG.

Importantly,  a natural analytical continuation of $S_{\bf w}$
to $d=4-2\epsilon$ dimensions exists and preserves its
Weyl gauge invariance. This is done using the Weyl gauge covariant
Weyl-Ricci scalar $\hat R$ as DR regulator,  instead of a  DR scale
$\mu$ which would break explicitly this symmetry if used.
The regularised Weyl gauge invariant action in $d\!=\! 4-\! 2\epsilon$ is
then just like $S_{\bf w}$ of (\ref{WWW}), but with the integrand multiplied by
$(\hat R^2)^{(d-4)/4}$.  The action is then Weyl gauge invariant in $d$ dimensions,
hence this {\it geometric} regularisation  ensures
there is no Weyl gauge anomaly \cite{DG1}. This is good news for the
quantum consistency of this gauge theory.

There exists a more fundamental  Weyl {\it gauge invariant} action
in arbitrary $d$ dimensions in Weyl geometry than action (\ref{WWW}), called
Weyl-Dirac-Born-Infeld action (WDBI) \cite{DBI,WDBI}
\bea\label{hh}
S_{\bf w}'\!\!\! & = &\!\!\!
\int d^d x \big\{-\det \big[a_0 \, \hat R\,g_{\mu\nu}+ a_1\, \hat R_{\mu\nu}
  +a_2\,\hat F_{\mu\nu}\big]\big\}^\frac{1}{2},
\eea
with dimensionless coefficients $a_0, a_1, a_2$.
This action is well-defined  and does {\it not}
need an ultraviolet (UV) regulator/regularisation (if one starts with its $d=4$ version,
all one has to do is a pure analytical continuation $d=4\ra d=4-2\epsilon$).
This is the only  theory we are aware of with this property.
Its leading order in a
series expansion in $\xi^2\!\sim\!\Lambda/M_p^2$ is shown below
with dimensionless physical  couplings $\xi, \eta, \alpha$ \cite{DBI,WDBI}
\be\label{hh2}
S_{\bf w}'=\!\int \!\! d^d x \sqrt{g}\,
\Big\{
\Big[\frac{1}{4!\,\xi^2}\hat R^2
 -\frac{1}{\eta^2} \hat C_{\mu\nu\rho\sigma}^2
 -\frac{1}{4\,\alpha^2}\, \hat F_{\mu\nu}^2+\hat G
 \Big]\,(\hat R^2)^{d/4-1} \!\! + \cO\Big(\frac{1}{\hat R^3}\Big)
\Big\}+ a_0^{d/2} \cO\Big(\frac{a_i^3}{a_0^3}\Big).
\ee

\medskip\noindent
For a suitable choice of $a_0, a_1, a_2$ in terms of  $\xi, \eta, \alpha$, 
the leading order of $S_{\bf w}^\prime$
recovers {\it  exactly} the  geometrically regularised
$S_{\bf w}$ of (\ref{WWW}) \cite{DBI}
for $1/a_0\!\sim\!\xi^{4/d}\!\sim \! (\Lambda/M_p^2)^{2/d}\!\ll\! 1$.

One can also include the SM operators under $\det[...]$,
in the $d=4$ WDBI action (\ref{hh});
again no UV regulator is necessary for either  gravity or SM action obtained in
the leading order - 
geometry takes care of that \cite{WDBI}.
Being gauge invariant in $d=4-2\epsilon$ dimensions, WDBI action (including SM) is
Weyl-anomaly free and a  quantum-consistent gauge theory of gravity
\cite{review,WDBI}. This ends our summary of Weyl geometry and its physics.

\section{Weyl gauge invariant dressed operators and action}
\subsection{Wilson line dressing, emergent non-commutativity}

Consider again initial Weyl gauge transformation (\ref{WGS})
extended to include the scalar field $\phi$
\medskip
\bea\label{WGS2}
 g_{\mu\nu}^\prime=\Sigma^2 \,g_{\mu\nu},\quad
 \w_\mu'=\w_\mu -  \partial_\mu\ln\Sigma, 
\quad
\sqrt{g'}=\Sigma^{d} \sqrt{g},\qquad
\phi^\prime=\Sigma^{\tilde q_\phi}\,\phi. \qquad
\eea

\medskip\noindent
The charge of $\phi$ is $\tilde q_\phi\!=-(d-2)/2$ (its inverse mass
dimension); $\ln\phi$ is the would-be-Goldstone of Weyl gauge symmetry,
of geometric origin.
By analogy to a U(1) gauge symmetry, for the gauged dilatation symmetry
we introduce the so-called ``dressed'' fields which are gauge invariant
composite operators. These were first introduced by Dirac \cite{Dirac2}
who pioneered the concept of  Faraday line (now called
Wilson line \cite{Wilson})
and by \cite{mandelstam,mandelstam2,terning},
(for a review on Wilson lines see e.g. \cite{faraji}).
Being gauge invariant, these operators are physical and can  describe 
observables\footnote{For a U(1) gauge theory, a dressed electron
describes the electron  with its surrounding cloud of photons.}.
The dressed fields' method  enables  one to avoid
using gauge potentials (and thus gauge fixing)
 and can realise naturally the spontaneous breaking of the
gauge symmetry \cite{Francois,Francois2}\footnote{There is no need for a 
dressed gauge field since $F_{\mu\nu}$ is already a gauge invariant  observable.
}. This Wilson line formalism, well-known  in
gauge theories of internal symmetries, is applied here for
gauged dilatations (space-time) symmetry.

With this observation,  we  introduce a Weyl gauge invariant (dressed)
metric, $g_{\mu\nu}^\star$, for Weyl gauge  theories (\ref{WWW}) and
(\ref{hh}). Let us explore the consequences of doing so.
Consider then a Wilson line $V(x)$ applied to Weyl gauge covariant
fields, so the dressed states (denoted with a star) become  Weyl
gauge invariant:
\bea\label{trg}
g^*_{\mu\nu}=V^2 \, g_{\mu\nu},
\qquad
\dphi=V^{\tilde q_\phi} \phi.
\eea
%
This definition can be applied
to  any tensor $T$ that transforms covariantly, of  Weyl charge $\tilde q_T$.
For the dressed fields to be Weyl gauge invariant,
$V(x)$ must depend linearly on $\omega_\mu$ to compensate the change under (\ref{WGS}) 
and has the form of a path integral\footnote{In general consider
$V(x)=\exp\big\{\int d^4y\, f^\mu(x,y) \,\omega_\mu(x)\big\}$.
The ``dressed'' fields are invariant under (\ref{WGS}) if
$V^\prime=\Sigma^{-1}\, V$ which is true provided that
$\partial_\mu f^\mu(x,y)=\delta^{(4)} (x-y)$
as seen after an integration by parts.
Thus $V$ is singular.  A possible choice for
$f_\mu(x,y)$ is $f^0=f^1=f^2=0$, $f^3=\theta(x^3-y^3) \delta(x^0-y^0) \delta(x^1-y^1)
\delta(x^2-y^2)$ giving for the integrand in $V$:
$\int_{-\infty}^{x^3} dy^3\, \omega_3(y)$. More generally
$V(x)=\exp\big\{\int_{-\infty}^x dy^\mu\, \omega_\mu(y)\big\}$ \cite{Dirac2},
used above.}
\bea\label{V}
V(x)=\exp\Big\{\int_{-\infty}^x \omega_\mu(y)\,  dy^\mu\, \Big\}.
\eea

\medskip\noindent
$V(x)$ and the dressed fields are then  path functionals, and to complete their definition
we detail the path of integration shortly. 
Explicitly, we have
(to make Weyl coupling explicit one rescales $\omega_\mu\!\ra\! \alpha\,\omega_\mu$)\footnote{
The possibility of using this new  metric  was previously noticed  in \cite{Ohanian}.}
\medskip
\bea\label{map}
g^*_{\mu\nu}(x)=g_{\mu\nu}\,\exp\Big\{2\int_{-\infty}^x dy^\mu\, \omega_\mu(y) \Big\},
\qquad
\dphi(x)=\phi\,\exp\Big\{- \frac{d-2}{2} \int_{-\infty}^x dy^\mu\, \omega_\mu(y)\, \Big\},
\eea

\medskip\noindent
with a {\it reference point} $x_0=-\infty$.
The interpretation of   $g^*_{\mu\nu}$ is that of a (charged) metric
surrounded by the cloud of Weyl gauge bosons.
Under (\ref{WGS}) these fields are Weyl gauge
invariant\footnote{
More generally, for a tensor of charge $\tilde q_T$ that transforms as
$T^{* \mu\nu...}_{\alpha\beta...} = V^{\tilde q_T} \, T^{\mu\nu...}_{\alpha\beta...}$, then
$T^{*\,\mu\nu...\, \prime}_{\,\,\alpha\beta...} = T^{*\, \mu\nu...}_{\,\,\alpha\beta....}$.
}
\bea\label{global-local}
\dgp_{\mu\nu}=g^*_{\mu\nu},
\qquad
\dphip=\dphi.
\eea
The rescaling of $g_{\mu\nu}$ (``ruler'')
is compensated at every point $x$ on the path of the integral,
by the Wilson line (connection) which
transports the ``ruler'', such that
$g_{\mu\nu}^*$ stays gauge invariant.

  This is path-history {\it (global)} dressing as opposed to a
{\it local} dressing of the metric that could have been done by $\phi$ alone
and giving instead a gauge invariant $g_{\mu\nu}\phi^2$.
In the limit  $F_{\mu\nu}=0$ i.e.  $\omega_\mu$ is ``pure gauge'', 
$\omega_\mu\sim \partial_\mu\ln\phi$, only the path-independent contribution
exists to the path integral, due to would-be-Goldstone $\ln\phi$, in which
case a local dressing is obtained.

Transformations
(\ref{trg}), (\ref{map}) have some similarity to
a {\it non-local, field-dependent} gauge transformation,
but note the new fields are composite.
There is no gauge fixing  condition (restriction) on these fields here.

As anticipated, dressed variables are   non-local  functionals of the path integral of
Wilson line. What is this path? It is the geodesic of Weyl gauge covariant (metric)
formulation ($\hat\nabla_\mu$),  defined by
$\tilde\Gamma_{\mu\nu}^\alpha$, the  so-called ``autoparallel''.
 $\tilde\Gamma$ is gauge invariant which is relevant for the invariance
of the dressed fields. The equation of the geodesic parametrised by
$y^\mu(\lambda)$ in Weyl geometry is, naturally, the Weyl gauge covariant version 
of that in Riemannian geometry, which means replacing $\nabla_\mu \ra \hat \nabla_\mu$.
This gives a Weyl geodesic \cite{Ligo}
\bea\label{Weyl-geo}
Y^\nu\hat\nabla_\nu Y^\mu=0,\qquad Y^\mu\equiv\frac{\partial y^\mu}{\partial\lambda}.
\eea
With $Y^\mu=e_a^\mu Y^a$ of Weyl charge (-1) given by $e_a^\mu$, then
$\hat\nabla_\nu Y^\mu=(\tilde\nabla_\nu -\omega_\nu)Y^\mu$.
Explicitly
\medskip
\bea\label{auto}
y^{\mu\,\prime\prime}(\lambda) +\tilde\Gamma_{\alpha\beta}^\mu \,\, y^{\alpha \,\prime}(\lambda)\,
y^{\beta \,\prime}(\lambda) -\omega_\alpha\, y^{\mu\,\prime}(\lambda) y^{\alpha \,\prime}(\lambda)=0.
\eea
%
where $y^{\alpha\, \prime}(\lambda)$ is the derivative of $y^\alpha$ with
respect to $\lambda$.
The path is a solution starting in the asymptotic region at $\lambda=0$,
 $y^\mu(0)=-\infty$ (where $\omega_\mu\ra 0$); the endpoint of 
Wilson line (\ref{V}) is  $x^\mu=y^\mu(\lambda=1)$.
Increasing $\lambda$ extends the history, the earlier path being unchanged.
Eq.(\ref{auto})  can be used  to perform the integral,
but this is not needed here (see Appendix~\ref{AC}).

There are important  implications for geometry and physics
when using dressed variables   $g_{\mu\nu}^*$, $\phi^*$ (instead of
$g_{\mu\nu}$,  $\phi$)\footnote{Note that these fields  are all of
 geometric origin, no matter is considered.}.
The first  consequence is that when acting on  dressed fields
\bea\label{com}
    [\partial_\mu, \partial_\nu]\, g^*_{\alpha\beta}=2\,\, F_{\mu\nu} g^*_{\alpha\beta},
    \qquad
    [\partial_\mu,\partial_\nu] \,\dphi=- \frac{d-2}{2} \,\, F_{\mu\nu}\,\dphi.
\eea

\medskip\noindent
where numerical factors are Weyl charges of $g_{\alpha\beta}$ and $\phi$;
$\partial_\mu\equiv \partial/\partial x^\mu$ is the partial derivative.

Let us detail:  a dressed field or $V(x)$ is a path functional -
treated as a   function of the end-point $x^\mu$ alone, it is ill-defined
for having different values for different paths ending at same $x^\mu$.
So one must clarify how the Wilson line path $y^\mu(\lambda)$
behaves when its end-point $x^\mu$ changes under the  action of
a partial ($\partial_\mu\equiv \partial/\partial x^\mu$) or of a total derivative ($d$).
For the action of a partial space-time derivative $\partial_\mu$
the argument is locality: the only possible deformation
of the path, consistent with the {\it  local} space-time derivative
$\partial_\mu$, takes place (locally) at the tip of the path
(in the neighbourhood of endpoint $x^\mu$)
with  path history held fixed, see Appendix~\ref{B}.
Therefore, we find  from (\ref{V})  $\partial_\mu V\! =\!\omega_\mu V$
and eq.(\ref{com}) follows (eq.(\ref{ac})).
This is known in gauge theories with  gauge {\it invariant}
Wilson-line-dressed fields since early works of  \cite{terning,mandelstam,mandelstam2}.
The situation is different for a {\it total} differentiation
of  Wilson line ($dV$) that  involves  non-local path-family
functional variation (non-local deformation), see eq.(\ref{B8}).
Relation (\ref{com}) states that if $F_{\mu\nu}\!\not=\! 0$,
$V(x)$ or $g_{\mu\nu}^*$, $\phi^*$ are not  functions of $x^\mu$ alone;
their path dependence and holonomy are  manifest  in (\ref{com});
scale holonomy is encoded { locally} at a point
in non-commutativity (\ref{com})  due\,to\,flux through infinitesimal
plaquette area $dx^\mu dx^\nu$ near endpoint $x$.

The relation $\partial_\mu V\! =\!\omega_\mu V$
is consistent with $\hat\nabla$-covariant constancy of Wilson line
in all directions (see later, eq.(\ref{gim}))
as generator of parallel transport, and with the
geometry-derived autoparallel (\ref{Weyl-geo})
generated by same $\hat\nabla$ that Weyl-covariantises  Riemannian geodesic,
see later.

There is a  physical interpretation why
path deformations under $\partial_\mu$ are local (at the tip of the path, path fixed),
based on causality:
the Wilson line is the transporter of the ruler (metric) along an actual history
that records changes along that history. Moving  forward in time
the endpoint extends the history along $y^\mu(\lambda$), 
but nothing about the past changes. In other words $\partial_0 V$ derivative
(at the tip of the path) compares its own extension with its history, not with
an alternative history of  different path (if the path were not fixed).
Causality supports this argument for time derivatives,
so $\partial_0 V=\omega_0 V$.  Lorentz covariance  extends
this  argument to spatial derivatives so $\partial_\mu V\!=\!\omega_\mu V$ and
(\ref{com})   follows.\footnote{
The relation $\partial_\mu V=\omega_\mu V$ (due to local path deformation)
is exactly the condition for the dressing to commute with 
covariant differentiation (Appendix~\ref{B}) that enables
Weyl-Riemann duality  map, see later.}

Non-commutativity (\ref{com}) induced by  Wilson line and
Weyl gauge symmetry at high scales, is present {\it only when acting on
 the vector space of  dressed fields!}
If there is no gauge symmetry e.g. $\w_\mu$ is not dynamical or
the symmetry is broken and  massive $\omega_\mu$ decouples  at low
energies  \cite{Ghilen0}, then $F_{\mu\nu}\!=\!0$, the commutators in
(\ref{com}) vanish and one restores commutativity in this space.
Unlike in other theories where it is assumed or added ad-hoc,
here non-commutativity is a consequence of using (gauge invariant)
dressed fields, which are path-dependent\footnote{
One can  check  Jacobi identities for $x^\rho$, $\partial_\mu$ operators
acting on dressed fields;
denoting  $[x^\mu, x^\nu]=\theta^{\mu\nu}$ and assuming usual
  $[x^\rho, \partial_\mu]=\delta^\rho_\mu$, these identities are respected
  provided $\theta^{\mu\nu}\!=$constant  or $\theta^{\mu\nu}=0$.
  For simplicity we restrict to the case  $\theta^{\mu\nu}=0$.
  In fact if $\theta^{\mu\nu}\!=$constant, expected to be $\propto (F^{-1})^{\mu\nu}$,
  the equation of motion for $\omega_\mu$ is not respected.
  More generally,  if $[x^\rho,\partial_\mu]=\delta_\mu^\rho$
is allowed to have corrections $\cO(1/m_\omega)$, a generalised uncertainty principle
\cite{mignemi} leads to non-commutative coordinates.
We do  not study this here.}.

A second consequence for geometry is that from (\ref{tGamma}), (\ref{LC})
\medskip
\bea\label{eq}\tilde\Gamma_{\mu\nu}^\rho(g,\omega)=
\tilde\Gamma^\rho_{\mu\nu}(\dg,0)=\Gamma_{\mu\nu}^\rho(\dg).
\eea

\medskip\noindent
Therefore,  Levi-Civita connection of the dressed metric, $\Gamma_{\mu\nu}^\rho(\dg)$,
is identical to  Weyl connection $\tilde\Gamma_{\mu\nu}^\rho$,
and  is Weyl gauge invariant, too, being function of $g_{\mu\nu}^*$.
Further, using the (local) action of partial derivative on $V(x)$, eq.(\ref{ac}),
that the covariant derivative is supposed to compensate it locally,
then we have ($V$  has charge $-1$)
\bea\label{gim}
\hat\nabla_\mu \exp\Big\{\int_{-\infty}^x\w_\alpha\,dy^\alpha\Big\}=0
\quad\Ra\quad
\hat\nabla_\mu g^*_{\alpha\beta}=0.
\eea

\medskip
Since $g^*_{\alpha\beta}$ has vanishing charge and using definition (\ref{qq}),
we now have a  Weyl gauge  invariant and  {\it metric} formulation in terms
of\footnote{
More generally, for a ``dressed''  tensor 
$\hat \nabla_\lambda T^{*\,\mu\nu...}_{\,\,\alpha\beta...}\equiv
\tilde\nabla_\lambda 
T^{*\,\mu\nu...}_{\,\,\alpha\beta...}
=\nabla_\lambda^* T^{*\, \mu\nu...}_{\,\,\alpha\beta...}.
$}$^,$\footnote{
Eq.(\ref{gim}) is also seen from
$
\tilde\nabla_\mu  g^*_{\alpha\beta}=
\partial_\mu g^*_{\alpha\beta}
-\tilde\Gamma_{\mu\alpha}^\rho\,g^*_{\rho\beta}
-\tilde\Gamma_{\mu\beta}^\rho\,g^*_{\rho\alpha}
=\exp\Big\{ 2\int_{-\infty}^x \omega_\mu dy^\mu\Big\}
  \big[ \tilde\nabla_\mu g_{\alpha\beta}+2 \omega_\mu \,g_{\alpha\beta}\big]
  =0
  $} $g_{\mu\nu}^*$ since:
\medskip
\bea\label{nablas}
{  \tilde\nabla_\mu g^*_{\alpha\beta}=0,
    \qquad
  \tilde\nabla_\mu\equiv \tilde\nabla_\mu(\tilde\Gamma(g,\omega))=
  \nabla_\mu (\Gamma(\dg))\equiv \nabla_\mu^*.
  }\eea

\medskip
To conclude,  we  have a  map (\ref{map}), (\ref{eq}) from Weyl geometry to a
(metric) Riemannian description defined by
$\nabla_\mu^*$      
and dressed metric $g_{\mu\nu}^*$, that is Weyl gauge  {\it invariant}; this is possible  at the cost
of  non-commutativity at high energy scales (UV).  Weyl connection (geometry)
becomes Riemannian connection (geometry) of path-dependent  $g_{\mu\nu}^*$.
Like Weyl geometry, the Riemannian picture depends not only on the (dressed) metric, but also
on $\omega_\mu$, since $F_{\mu\nu}$ is controlling  non-commutativity.
If $\omega_\mu$ becomes massive and decouples, $F_{\mu\nu}=0$ and commutativity is restored
at low scales.

Section~\ref{2} showed how  vectorial non-metricity ($\omega_\mu$)
of the rhs of (\ref{conn}) was moved into $\tilde\nabla_\mu$ (giving $\hat\nabla_\mu$),
to obtain the Weyl gauge covariant metric formulation,
$\hat\nabla_\mu g_{\alpha\beta}\!=\!0$.
Alternatively, moving  $\omega_\mu$ of the rhs of (\ref{conn})
in  $g^*_{\alpha\beta}$ as  here, gives eq.($\ref{nablas})$ and so,
the initial affine, non-metric formulation $\tilde\nabla_\mu (\tilde\Gamma)$
has now become {\it metric and gauge invariant}.
This is trading a non-metric Weyl connection ($\tilde\nabla_\mu(\tilde\Gamma))$
for a  gauge-invariant metric geometry with non-commuting derivatives
(on dressed fields).
This is another way to regard this Riemannian
description of dressed fields (see also eq.(\ref{curvature-1p})).

\subsection{Dressed operators}

Consider next the dressed curvature tensors. From 
(\ref{curvature-1}) with (\ref{eq})
\medskip
\bea
\tilde R^\rho{}_\sigma{}_{\mu \nu}
=
 \partial_\mu  \Gamma^\rho_{\nu \sigma}(g^*)
  - \partial_\nu \Gamma^\rho_{\mu \sigma}(g^*) +  \Gamma^\rho_{\mu \tau}(g^*)\,
   \Gamma^\tau_{\nu \sigma}(g^*) -  \Gamma^\rho_{\nu \tau} (g^*) \,\Gamma^\tau_{\mu \sigma}(g^*)
   \equiv R^\rho{}_\sigma{}_{\mu \nu} \Big\vert_{g^*} 
  \label{curvature-1p}
\eea

\medskip\noindent
In the last step we used the definition of Riemann tensor  in Riemannian geometry,
denoted $R^\rho_{\,\,\,\sigma\mu\nu}$. A subscript $g^*$ indicates the use of Levi-Civita
connection of dressed  $g_{\mu\nu}^*$, eq.(\ref{map}).
Also $R_{\sigma\nu}=R^\rho_{\,\,\sigma\rho\nu}$ and $R=g^{\sigma\nu} R_{\sigma\nu}$ of
Riemannian geometry. We denote
\medskip
\bea
R^\rho_{\,\,\,\sigma\mu\nu}(g^*)\equiv R^\rho_{\,\,\,\sigma\mu\nu}\Big\vert_{g^*},
  \qquad
R_{\sigma\nu}(g^*)\equiv R_{\sigma\nu}\Big\vert_{g^*},
\qquad
R(g^*)\equiv R\Big\vert_{g^*}.
\eea

\medskip
Relation (\ref{t1}) acquires now a Riemannian ``look''
\medskip
\bea
[\nabla_\mu^*,\nabla_\nu^*]\, T^\rho=R^\rho_{\,\,\,\sigma\mu\nu}(g^*)\,T^\sigma.
\eea

\medskip\noindent
Contractions, raising and lowering
of these curvature tensors indices are then made by $g_{\mu\nu}^*$.
The dressed Riemann tensor  has  Bianchi identities
 consequence of those in the Weyl covariant formulation
(for $\hat R^\mu_{\,\,\,\nu\rho\sigma}$ and $\hat F_{\mu\nu}$)\footnote{The notation
$[\lambda\rho\sigma]$ stands for cyclic permutations of the indices inside
the brackets.} and  re-written below with a Riemannian ``look'':
 \medskip
 \bea
\nabla_{[\lambda}^* R^\mu_{\,\,\,\,\vert\nu\vert\rho\sigma]}(g^*)=0,
\qquad\qquad
R^\mu_{\,\,\,[\nu\rho\sigma]}(g^*)=0.
\eea

\medskip
Next, using
(\ref{relations-curvatures}), (\ref{curvature-1p}) and Appendix~\ref{A}, 
the following relations hold true:
\medskip
\bea\label{rel}
\hat R^\rho_{\,\,\,\sigma\mu\nu}&=&
 R^\rho_{\,\,\,\sigma\mu\nu}(g^*)-\delta^\rho_\sigma \,F_{\mu\nu},
\\[5pt]
\hat R_{\sigma\nu}&=&R_{\sigma\nu}(g^*)-F_{\sigma\nu},\label{r2}
\\[5pt]
\hat R & = & R(\dg)\,V(x)^2.
\label{rl}
\eea

\medskip\noindent
Regarding the path of integration in $V(x)$ in (\ref{rl}),
consider the geodesic 
\medskip
\bea\label{Geo}
Y^{* \nu} \nabla_\nu^* Y^{* \mu}=0,
\eea

\medskip\noindent
computed with gauge invariant  $g_{\mu\nu}^*$, with
$Y^*_{ \mu}\equiv  e_\mu^{a *} Y_a$ and $Y^{*\mu}\equiv e^{* \mu}_a Y^a$.
This is exactly the familiar Riemannian geodesic
$Y^\nu \nabla_\nu Y^\mu=0$, of Riemannian $\nabla_\mu$,  written
in terms of  $g_{\mu\nu}^*$ regarded as independent variable.
If in $Y^{* \nu} \nabla_\nu^* Y^{* \mu}=0$  we use (\ref{map}) for
the  metric/vielbein, then use (\ref{gim}), (\ref{nablas}),
we find exactly the Weyl geodesic equation of (\ref{Weyl-geo}).
Hence the geodesic in the Weyl covariant (metric) formulation 
becomes in  Riemannian geometry the familiar Riemannian
geodesic, but evaluated for $g_{\mu\nu}^*$ 
 (in $V(x)$ in  (\ref{rl}) we assume this is the case).

\subsection{Weyl gauge invariant action of dressed fields}

Using (\ref{rel}) to (\ref{rl}) and Appendix~\ref{A}, we have in $d$ dimensions
\bea\label{first}
\sqrt{g}\,\hat R_{\sigma\nu}^2& =& \sqrt{\dg}\,\big[ R^2_{\sigma\nu}(\dg)
  -(d-1) \, F_{\mu\nu}(g^*)^2\big]\,V(x)^{4-d}
\\[9pt]
\sqrt{g} \, \hat R^2 &=& \sqrt{\dg}\,R^2(\dg) \,V(x)^{4-d}
\\[9pt]
\sqrt{g}\, \hat F_{\mu\nu}^2 &=& \sqrt{\dg}\, F_{\mu\nu}^2(\dg)\,V(x)^{4-d}
\\[9pt]
\sqrt{g}\, \hat G &=& \sqrt{g^*}\, \big[ G(g^*) + c_1\, F_{\mu\nu}(g^*)^2 \big]\, V(x)^{4-d},
\label{GG}\\[9pt]
\sqrt{g}\,\hat C^{\rho\,2}_{\,\,\,\mu\nu\sigma}&=& 
\sqrt{g^*}\,\, \big[C^{\rho\,2}_{\,\,\,\mu\nu\sigma}(g^*)+c_2\,F_{\mu\nu}(g^*)^2\big]
\label{last}
\eea

\medskip\noindent
All contractions of indices in the rhs are done by
 $g_{\mu\nu}^*$, $G(g^*)$ is the Euler-Gauss-Bonnet term in Riemannian geometry
evaluated for $\Gamma(g^*)$;
$c_1=2 \, d^2- 7 d+8,\,\,c_2=(d^2-2 d+4)/(2-d)$.

Consider now the Weyl quadratic action of WG in the  covariant formulation in $d=4$, 
 eq.(\ref{WWW}), which we re-write here with coefficients
$\alpha_i$ ($i=1,..4$) that can be read from (\ref{WWW})
\medskip
\bea\label{rs}
S_{\bf w}=\int d^4x\,\sqrt{g}\,
\Big[ \alpha_1 \hat R^2 + \alpha_2 \,\hat C_{\mu\nu\rho\sigma}^2 + \alpha_3\,
  \hat F_{\mu\nu}^2 +\alpha_4\,\hat G\Big].
\eea

\medskip\noindent
This can be written in  Riemannian geometry
defined by $\Gamma(g^*)$, using eqs.(\ref{first}) to (\ref{last})
\medskip
\bea\label{RS}
\qquad\qquad
S_{\bf w}=\int d^4x\,\sqrt{g^*}\,
\Big[ \alpha_1  R^2 + \alpha_2 \, C_{\mu\nu\rho\sigma}^2 + \tilde\alpha_3\,
  F_{\mu\nu}^2 +\alpha_4\, G\Big]\Big\vert_{g^*}.\qquad\qquad 
\eea

\medskip\noindent
This action has a similar form to (\ref{rs}) up to  a redefinition: $\alpha_3\ra
\tilde\alpha_3=\alpha_3+ c_2\,\alpha_2+c_1\,\alpha_4$.
Eq.(\ref{RS}) is also Weyl gauge invariant and  metric with respect to
$g^*_{\alpha\beta}$ but we have non-commutativity in the sense that partial derivatives
acting on $g_{\mu\nu}^*$ do not commute, see (\ref{com}).
It is also {\it non-local} since the dressed operators are so, as discussed.

Conversely,  we can start from  a Riemannian geometry action that is (globally)
scale invariant,
as in (\ref{RS}), evaluated for some metric $g_{\mu\nu}^*$ and do a transformation
$g_{\mu\nu}^*\ra V(x)^2 g_{\mu\nu}$ with the derivative (non-holonomic) constraint
$\partial_\mu\ln V(x)=\omega_\mu$.
Under this transformation, we have
$R(g^*)\ra V(x)^{-2} \,[R- 6\nabla_\mu\nabla^\mu \ln V
  - 6 (\nabla_\mu \ln V) (\nabla^\mu \ln V)]$
so 
$R(g^*)\ra V(x)^{-2} (R- 6\nabla_\mu\omega^\mu -6 \omega_\lambda \omega^\lambda)
=V(x)^{-2} \hat R$ and
$\sqrt{g^*}\ra \sqrt{g}\,V^{4}$. Using this and that in $d=4$
the gauge kinetic term $\sqrt{g} F_{\mu\nu}^2$ is
invariant under rescaling  the metric, while $G$ is a total derivative,
we find exactly  action (\ref{rs}).

The equivalent description Weyl vs Riemann that we saw for $d=4$
extends to {\it regularised} actions
in  $d$ dimensions
\bea\label{e1}
S_{\bf w, d}&=&
\int d^d x\,\sqrt{g}\,
\Big[ \alpha_1 \hat R^2 + \alpha_2 \,\hat C_{\mu\nu\rho\sigma}^2 + \alpha_3\,
  \hat F_{\mu\nu}^2 +\alpha_4\,\hat G\Big] \,(\hat R^2)^{(d-4)/4}
\\
&=&\int d^d x\,\sqrt{g^*}\,
\Big[ \alpha_1  R^2 + \alpha_2 \, C_{\mu\nu\rho\sigma}^2 + \tilde\alpha_3\,
  F_{\mu\nu}^2 +\alpha_4\, G\Big]\,(R^2)^{(d-4)/4}\Big\vert_{g^*}. 
\label{e2}\eea

\medskip\noindent
The first line is in Weyl geometry notation, while the second line is the same action
in a Riemannian notation, for the dressed metric $g_{\mu\nu}^*$. This gives two mathematically
equivalent descriptions,  {\it both regularised and Weyl gauge invariant.}

The equivalent descriptions in   Weyl  vs 
Riemannian geometry  also apply to
the more fundamental  WDBI gauge theory action. 
This  is  seen using  eqs.(\ref{rel}) to (\ref{rl}),
and metric (\ref{map}).
Indeed,  $S_{\bf w}'$ of (\ref{hh}) becomes in Riemannian geometry
of $\Gamma(g^*)$
\bea\label{swp2}
S_{\bf w}'= \int d^dx \,\big\{-\det \big[ a_0 \, R \, g_{\mu\nu}^*
  +a_1 R_{\mu\nu} +\tilde a_2 F_{\mu\nu}\big]\big\}^{1/2}\Big\vert_{g^*}.
\eea

\medskip\noindent
up to a redefinition of (dimensionless)  $a_2\ra \tilde a_2$.
$S_{\bf w}^\prime$  is Weyl gauge invariant in $d$ dimensions.

We thus have equivalent descriptions of the action, in
Weyl and in Riemannian geometries, the latter in terms of gauge invariant
$g^*_{\mu\nu}$. Either description can be used in applications.

\subsection{Equations of motion:  Weyl vs Riemann}\label{3.4}

The duality map of similar actions in Weyl vs Riemann geometry can
be extended to the equations of motion, but there is a path-dependent ``memory''
effect missing on the Riemannian side, as we see shortly.
Consider action (\ref{rs}) in Weyl geometry in $d\!=\!4$ for $\alpha_2\!=\!0$\footnote{
The more general case of an action that includes the
additional term $\hat C^{2}_{\mu\nu\rho\sigma}$ in
(\ref{rs}) is immediate using \cite{CDA2}.}
\bea \label{fwg}
S_{\bf w}=\int d^4x\sqrt{g}\,\, \big[\alpha_1 \,\hat R^2+ \alpha_3 \,\hat F_{\mu\nu}^2\big]
=\int d^4x \sqrt{g}\,\, \big\{-\alpha_1 \big[\,2\phi^2\,\hat R+\phi^4\big]
+\alpha_3 \hat F_{\mu\nu}^2\big\}.
\\
\nonumber\eea

\medskip\noindent
In the second step we gave an equivalent linearised version of the action by
using the ``dilaton'' field $\ln\phi$, as explained in the text after eq.(\ref{WWW}),
so the two actions above are classically equivalent, and we use the second one below.
Then the equation of motion of $g_{\mu\nu}$ and $\omega_\mu$ are
\medskip
\bea
\phi^2 \Big[ \frac12 \Big(\hat R +\frac12\,\phi^2\Big)  g_{\mu\nu}- \hat R_{(\mu\nu)}\big]
+ \Big(\hat \nabla_{(\mu}   \hat\nabla_{\nu)}  \phi^2-g_{\mu\nu}\hat\Box\phi^2\Big)
+ \frac{\alpha_3}{\alpha_1}\, T_{\mu\nu}=0,
\\[5pt]
\hat\nabla_\mu \phi^2+\frac{\alpha_3}{3\,\alpha_1}\hat\nabla_\nu F_{\alpha\mu} g^{\alpha\nu}=0,
\eea
where
\medskip
\bea
T_{\mu\nu}= \hat F_{\mu\alpha} \hat F_{\nu\beta} g^{\alpha\beta}
  -\frac14 g_{\mu\nu} \hat F_{\rho\sigma} \hat F_{\kappa\delta} g^{\rho\kappa} g^{\sigma\delta}.
\eea

\medskip\noindent
Multiply the above equations by $\exp(-2\int^x \omega_\alpha dx^\alpha)$
and use eqs.(\ref{map}), (\ref{gim}), (\ref{nablas}),
(\ref{rel}) to (\ref{rl}) and that  dressed fields have no charge.
The equations of motion become $\dphi^2=-R(\dg)$ and
\medskip
\bea\label{m*}
\dphi^2 \Big[ \frac12\,\Big(R(\dg) +\frac12\,\dphi^2\Big)  g^*_{\mu\nu}\!
  -  R_{(\mu\nu)}(\dg)\Big]
\!+\! \Big[ \nabla^*_{(\mu}   \nabla^*_{\nu)}  \dphi^2
  \!-g^*_{\mu\nu}\, \dg^{\alpha\beta}\,
  \nabla^*_\alpha\nabla^*_\beta\dphi^2\Big]
\!\!+\! \frac{\alpha_3}{\alpha_1} T^*_{\mu\nu}\!\!\!\!
&=&
\!\!\!0, \qquad
\\[5pt]
\nabla_\mu^* \dphi^2+\frac{\alpha_3}{3\,\alpha_1}\nabla_\nu^* \hat F_{\alpha\mu} g^{*\,\alpha\nu}=0,
\label{m**}\eea
where
\medskip
\bea
T^*_{\mu\nu}=T_{\mu\nu} \exp\Big\{-2\int^x_{-\infty}\omega_\alpha dx^\alpha\Big\}=
\hat F_{\mu\alpha} \hat F_{\nu\beta} g^{*\,\alpha\beta}
-\frac14 g^*_{\mu\nu} \hat F_{\rho\sigma} \hat F_{\kappa\delta} g^{*\,\rho\kappa} g^{\,*\sigma\delta}.
\eea

\medskip\noindent
Above we used that $\hat R_{(\mu\nu)}
=\tilde R_{(\mu\nu)}=R_{(\mu\nu)}(\dg)$, see (\ref{relations-curvatures}) and (\ref{r2}).

Next consider an action of structure similar to (\ref{fwg}) but in Riemannian geometry
\medskip\bea\label{S_R}
S_{\bf R}&=&
\int d^4x\sqrt{g}\,\, \big[\alpha_1 \, R^2+ \alpha_3 \,F_{\mu\nu} F_{\alpha\beta}\,
  g^{\mu\alpha} g^{\nu\beta}\big]
\nonumber\\
&=&\int d^4x \sqrt{g}\,\, \big\{-\alpha_1 \big[\,2\phi^2\, R+\phi^4\big] +\alpha_3 F_{\mu\nu}
  \, F_{\alpha\beta}\,   g^{\mu\alpha} g^{\nu\beta}
  \big\},
\eea

\medskip\noindent
where we linearised the quadratic term as before;
$F_{\mu\nu}$  is as in Weyl geometry,
$F_{\mu\nu}=\partial_\mu\omega_\nu-\partial_\nu\omega_\mu=\hat F_{\mu\nu}$.
If in $S_{\bf R}$ we replace $g\ra g^*$ we find  exactly 
action (\ref{fwg})\footnote{This
is a particular case of the equivalence of actions
(\ref{rs}), (\ref{RS}) for $\alpha_2=0$; $G$ is a total derivative.}
that lead to eq.(\ref{m*}).

Suppose now one first writes the equations of motion
for independent variable $g_{\mu\nu}$ for action (\ref{S_R});
 {\it after that} replace in these equations all fields by their
``dressed'' version i.e. replace
$g_{\mu\nu}\ra g_{\mu\nu}^*$, $\phi\ra\dphi$  and $\Gamma(g)\ra \Gamma(g^*)$
and $\nabla_\mu\ra\nabla_\mu^*$, $\hat R\ra R(g^*)$, $R_{\mu\nu}\ra R_{\mu\nu}(g^*)$, etc.
The result obtained is  exactly that shown by equation (\ref{m*}) for the metric.
Similar conclusion  for $\phi$.
Therefore,  the  map of the two geometries, $g_{\mu\nu}\ra g_{\mu\nu}^*$, 
also applies at the level of equations of motion for $g_{\mu\nu}$ and $\phi$.
In  (\ref{m*}) the two partial derivatives acting on the dressed metric
do {\it not} commute anymore\footnote{
This is traced back to the previous derivative (non-holonomic) character of
the constraint $\partial_\mu\ln V(x)=\omega_\mu$.}, but since $g_{\mu\nu}$ is symmetric,
so is their action and then the classical equation of motion for the metric
does not ``feel'' non-commutativity induced by  the anti-symmetric part.
Thus, starting from the Riemannian picture,  dressing of the metric and
of $\phi$ commutes with computing their equations of motion from this action.
This is because  dressing commutes with 
  covariant differentiation: $(\hat \nabla_\mu \phi)^*\!=\!\hat\nabla_\mu\phi^*$
and  similar for the metric.

The situation is different for $\omega_\mu$. 
If starting from (\ref{S_R}) we write the equation of motion of $\omega_\mu$
and $g_{\mu\nu}$ and then replace them by their dressed values,
then eq.(\ref{m**}) is not recovered;
the first term in (\ref{m**}), which is the Weyl gauge current \cite{Ghilen0,SMW},
is  missed by Riemannian action (\ref{S_R}).
This term  arises through the variation that included
path-history dependence (see Appendix~\ref{B}),
and is not captured by this approach.  In other words,
the dressing of the metric does not ``commute'' with the derivation of
the equation of motion for $\omega_\mu$. Obtaining
the correct equation of motion for $\omega_\mu$
requires first including the path-history dependence of the metric i.e.
using $g_{\mu\nu}^*$. That $\omega_\mu$ requires respecting the ``history''
is not  surprising: its field strength controlls non-commutative
effects and scale holonomy information\footnote{See also the previous footnote.
Note also that if $\omega_\mu$ were pure gauge, then $F_{\mu\nu}=0$,
the commutator of partial derivatives would vanish and there is no
path-history dependence. But then there is no Weyl conformal geometry,
no Weyl gauge symmetry and no $\omega_\mu$ in the spectrum.}.
This is the underlying difference between Weyl and Riemannian
geometry when describing the Weyl gauge invariant phase
at the level of equations of motion. 
 Non-commutativity  (\ref{com}) is also seen during  canonical
 quantisation of the theory.

 Finally, we give the exact details of the map of the metric between
 Weyl and Riemann geometries for action (\ref{fwg}).
 Assume a static case  with spherical symmetry, with $\omega_{2,3}=0$, and
 use that $\omega_1=\partial_1 \ln \phi$ from  equations of motion,
 (with spherical coordinates notation $x^\mu =(t,r,\theta,\varphi)$).
 The Wilson line is in this case controlled by the
 temporal mode quantum fluctuation $\omega_{0}$ that has a non-zero $F_{\mu\nu}$
 (we assume a background $\omega_\mu=0$).
Then the dressed metric has the Weyl gauge invariant form
\medskip
\bea\label{wd}
g_{\mu\nu}^*(x)=g_{\mu\nu}(x) \phi^2(x)\,\exp\Big\{2\int_{-\infty}^x 
\omega_0\, dt \Big\}.
\eea

\medskip\noindent
The contribution of the  mode $\omega_1=\partial_1\ln\phi$ to the dressing
 is independent of the path of
 integration and gives  the overall $\phi^2$ factor in (\ref{wd}).
 We  see again  that {\it local} dressing, by $\phi^2$ alone, is
 a particular case of Wilson line dressing when ignoring the
 path-history (possible  if $\omega_\mu$ is ``pure gauge'' i.e.
 when there is no true Weyl gauge symmetry and no Weyl geometry).
 One can compute integral (\ref{wd}) in various limits of interest,
 using the equation of motion for action (\ref{fwg})
 together with the path of integration parametrised as $t=t(\lambda)$, $r=r(\lambda)$,
 detailed in Appendix~\ref{AC}.
This expression is also valid for  the full quadratic action in $d=4$, eq.(\ref{rs})
which includes the  term $\hat C_{\mu\nu\rho\sigma}^2$, since this term does not affect the
equation of motion of $\omega_\mu$.

While the Wilson line factor is a classical geometric object,
it integrates quantum fluctuations of  $\omega_\mu$;
in this sense, the map $g_{\mu\nu}\ra g_{\mu\nu}^*$
between the two Weyl gauge invariant actions in Weyl vs Riemann
geometries ``knows'' about (non-local)
quantum effects, and then so does non-commutativity induced by
the (derivatives of) Wilson line.

\section{Physical implications}\label{s4}

Using dressed fields we found a new, Riemannian perspective on Weyl gauge symmetry
that defines Weyl  geometry:

\bigskip
\noindent
    {\bf (a)} {\bf  Non-local, duality map:}
    Weyl gauge symmetry  has an alternative
    description from a Riemannian view (of dressed fields):
Weyl connection is the Levi-Civita connection of the dressed metric $\Gamma(g^*)$
and $\hat \nabla_\mu$ is replaced by $ \nabla^*_\mu$.
This gives  an equivalent Riemannian formulation of connection $\Gamma(g^*)$
with $\nabla_\lambda^* g^*_{\mu\nu}=0$. So Weyl geometry
has an  equivalent (dual) description as Riemannian geometry of Weyl gauge invariant
dressed (physical) fields, at the cost of non-commutativity of $\partial$'s
when acting on these fields. These fields are non-local due to
Wilson line integral, and this is also true for the map
$g_{\mu\nu}\ra g_{\mu\nu}^*$. Non-commutativity  ($\propto F_{\mu\nu}$)
is  not added by hand, it is {\it emergent in the vector space
of Weyl gauge invariant fields/observables only} (eq.(\ref{com})), and is
induced by Weyl gauge symmetry!

\bigskip
\noindent
  {\bf (b)} {\bf  A physical scale from Weyl flux: }
In Weyl gauge theories all mass scales are generated by Weyl gauge symmetry breaking,
so Planck mass, $m_\omega$ and $\Lambda$ are set by $\langle\phi\rangle$ up to
couplings $\xi, \alpha$, see (\ref{scales}).
In the dressed picture the relevant vev is Weyl {\it gauge-invariant}
$\langle\phi^*\rangle$ so these scales are indeed {\it physical}.
The largest physical mass is $M_p$ (or $m_\omega\sim \alpha M_p$), so
$1/M_p$ is the shortest physical length of the theory.
Further, since dressed fields are non-local, so are their interactions which are spread
along Wilson-line paths and thus are not point-like. In the perturbative Weyl quadratic
gravity action this can lead to a softer UV behaviour, and, possibly
test the shortest length,
as in perturbative string theory, where the extended, non-local
nature of interactions gives a resolution bound\footnote{
A possible minimal length/area might be derived from an uncertainty
relation following from (\ref{com}).}.
The WDBI action follows a different, non-perturbative
logic, being more like a D-brane effective action,
whose leading order is the geometrically regularised
Weyl quadratic gravity. The same caveat as in string then applies: a resolution bound
in perturbative regime need not hold non-perturbatively, since D-branes can probe
distances smaller than string length. For WDBI, $1/M_p$ should then be understood
as a length scale where 
perturbative description changes \cite{DBI,WDBI},
not as a bound or resolution, since WDBI needs no regulator, see {\bf  (d)}.

\bigskip
\noindent
     {\bf (c)} {\bf Dual actions:}
The  quadratic   action in Weyl geometry  has a similar, equivalent
expression  in  Riemannian geometry of
dressed metric, at the cost of    non-commutativity in the UV.
This is also true for the more general  WDBI action.
The action on Riemannian side is Weyl gauge invariant since it is
a function of $g^*_{\mu\nu}$.
While the Wilson line is an otherwise (classical) geometric object,
with a background value $\overline\omega_\mu=0$ to respect Lorentz symmetry,
the dressed metric $g^*_{\mu\nu}$ and the Wilson line ($\sim\exp\int^x \omega_\mu dy^\mu$)
integrate {\it quantum} fluctuations of $\omega_\mu$ and their non-locality is thus of
quantum nature; the two Weyl gauge invariant actions in  Weyl and Riemannian
geometry are connected by a quantum effect.

\bigskip
\noindent
    {\bf (d)} {\bf  No UV regulator in WDBI:}
    The similar expression of the actions in Weyl vs
Riemannian geometry is also valid in arbitrary $d$ dimensions while preserving 
Weyl gauge symmetry, see eqs.(\ref{e1}), (\ref{e2}).
In particular, this is true  for $d=4-2\epsilon$,
$\epsilon\!\ll\! 1$. It is geometry itself
by its Weyl scalar curvature $(\hat R^2)^{-\epsilon/2}$ or
$(R^2(\dg))^{-\epsilon/2}$  in Riemannian case of dressed fields, that
gives the natural regulator\footnote{The presence of a scale as regulator
would break explicitly Weyl gauge symmetry.}  and a Weyl gauge invariant regularisation.
This is  endorsed mathematically by the WDBI action
\cite{DBI,WDBI} where {\it no } UV regulator is needed, see (\ref{hh}),
(\ref{swp2}),  and of which the regularised  Weyl quadratic gravity
is the leading order, see (\ref{hh2}). This   remains true if SM
is added on top;  geometry enforces a mathematically well
defined action~\cite{WDBI}. 
Also, as already mentioned, non-locality of dressed fields and thus of their interaction
in action (\ref{swp2})
with similarity to string interactions, suggests an improved UV behaviour in this case.

\bigskip
\noindent
    {\bf (e)\,}
  {\bf   Dressed fields as observables:}
As mentioned, it is the dressed
$\phi^*$ that is Weyl gauge invariant,  so
masses are generated by $\langle\phi^*\rangle$
(not by $\langle\phi\rangle$) and its Wilson line dressing
sums up quantum fluctuations  of $\omega_\mu$ along the integral path.
 Therefore  masses are in fact quantum corrected by $\omega_\mu$ fluctuations
 (in the Riemannian picture of dressed fields/metric). The ``gauge-history''
 dependence  of the Wilson line is compensated at every point (locally)
 by the transformation
 of $\phi$ itself, eq.(\ref{map}), such that $\phi^*$ and observable
 masses are gauge invariant.
 So  masses and dressed fields  include and carry a (global, non-local)
 result of integrated  {\it quantum} gauge-history of  the Wilson line!
 At low energies, dressing becomes irrelevant since massive $\omega_\mu$
decouples ($F_{\mu\nu}\!=\!0$) and Riemannian picture of dressed fields
recovers that of a  broken phase of ordinary, undressed fields (strictly
speaking, $[\partial_\mu,\partial_\nu]$ never vanishes {\it exactly}, but
is controlled by $\alpha\sim m_\omega/M_p$). In this broken phase one
recovers  the Einstein-Hilbert action \cite{Ghilen0}.

If one is using this broken phase to compute UV solutions for ordinary $g_{\mu\nu}$,
one recovers Schwarzschild-like solutions \cite{Harko2}. The singularity
of these solutions appears to be an artefact of a description that is using
the wrong (undressed) fields and metric since these are not gauge invariant
(but covariant!);
the metric singularity is then expected to be a gauge artefact. But a
singularity in (Weyl gauge invariant)
$g_{\mu\nu}^*$, if found,  would actually be a physical one\footnote{subject to
respecting diffeomorphism invariance.}. 
A well-defined action (\ref{swp2}) with $\det >0$
restricts $F_{\mu\nu}$ (relative to curvature terms
$\hat R_{\mu\nu}^*$, $R^* g_{\mu\nu}^*$), as in Born-Infeld electrodynamics.
This is expected to oppose a gravitational collapse into a point-like singularity
and possibly lead to a bounce,
geodesic completeness and no information loss.
This is a conjecture that must be tested on a solution for $g_{\mu\nu}^*$
of WDBI action.

\medskip
\noindent
{\bf (f)\,} {\bf ``Effective'' non-locality:\,}
It is interesting to compare the Weyl geometry perspective to the
Riemann view of dressed fields. 
At the deepest, fundamental level, we have a description that is a
{\it local, continuous, Weyl gauge invariant
quantum theory}, Weyl-anomaly free and with no masses,
with an underlying Weyl conformal geometry. 
In the traditional Riemannian view of ordinary (undressed) fields we
see  the low-energy limit of this theory, which is
a {\it broken} phase of Weyl gauge symmetry, where mass is generated.
This effective theory view is incomplete, hence it  may be regarded as a projective
representation or gauged fixed version of Weyl geometry picture \cite{Ghilen0}.

To avoid gauge-fixing artefacts, one is forced to use instead a Riemannian
picture  of Weyl gauge invariant dressed fields and action, supposed to give
something closer to an isomorphic representation of the fundamental, local theory
 in Weyl geometry.
 The (gauge-induced) quantum non-locality of the dressed fields, eqs.(\ref{map}),
 and non-commutativity, eq.(\ref{com}),   that appear in this dual
Riemannian picture,  are representation artefacts and the ``tax''  paid
to  translate Weyl geometry and its {\it Weyl gauge covariant fields} into 
Riemannian geometry  in which we live  of  {\it Weyl gauge invariant} 
observables like  masses that a (Riemannian) physical observer is measuring.
This  is an ``effective'' non-locality due to Weyl gauge invariance
and is a physical (!) artefact (in  Weyl geometry perspective) of using the
``wrong'' geometry, but the fundamental theory in Weyl geometry is local,
as stated\footnote{In other words, the ``tool'' is the problem, ``reality''
at the deepest layer  is ultimately local.}.
If we thus observe such ``effective'' quantum non-locality in our-world
Riemannian geometry  of Weyl gauge-invariant (dressed) fields, this is {\it not}
a concern, but  is evidence of a Weyl gauge symmetric phase
that integrates quantum fluctuations of $\omega_\mu$ (\ref{map})!
Quantum non-locality may be evidence of Weyl gauge symmetry!

\section{Entanglement in quantum non-local geometry}

{
  As an application,
we explore what Weyl gauge invariance and quantum non-locality
of the metric $g_{\mu\nu}^*$ may  imply for two entangled photons.
In  quantum mechanics this   refers to entangled {\it states} in Hilbert space 
and it leads to so-called  ``spooky'' non-local connection between particles
in flat  spacetime,  regardless of the distance.
A  detailed study of this non-local connection in curved space-time
is beyond the purpose of this work;  here we simply outline  a conceptually new
view that may  explain such non-local connection as a manifestation of
 quantum non-local Riemannian geometry of dressed metric!
}

In our-world Riemannian approach of  gauge-invariant observables,
described by dressed fields and metric $g_{\mu\nu}^*$, consider
the photon action,
$(-1 /4 \alpha^2) \int d^4x\,\sqrt{g^*} g^{*\,\mu\nu} g^{*\,\alpha\beta} \,\cF_{\mu\alpha} \,\cF_{\nu\beta}$
which can be re-written as $(-1/2)\int d^4x\, \sqrt{g^*} g^{*\,\mu\alpha} A_{\mu}\,
\cO_{\alpha\beta}^* \,A_\nu \, g^{*\,\beta\nu}$ ($A_\mu$ is not dressed). 
The photon propagator $\cA^*_{\mu\nu}(x_1,x_c)$ in Feynman gauge is then a solution to 
\bea\label{dme}
\cO_{\alpha\beta}^* \,\cA^{*\,\beta\nu}(x_1,x_c) =\frac{\delta_\alpha^\nu\,\delta^{(4)}(x_1-x_c)}{
  g^{*\, 1/4}(x_1) \, \, {g^{*\, 1/4}(x_c)}},\qquad
\cO_{\alpha\beta}^*=g_{\alpha\beta}^* \,\Box^*-R_{(\alpha\beta)}^*.
\eea
%
By comparing to the action without dressing (local, Riemannian case),
one sees that $\cO_{\mu\nu}^*=\cO_{\mu\nu}$ with $\cO_{\mu\nu}$ in
the absence of dressing, hence
 $\cA_{\mu\nu}^*(x_1,x_c)=\cA_{\mu\nu}(x_1,x_c)$, but:
$\cA^{*\,\mu\nu}(x_1,x_c)= V^{-2}(x_1) V^{-2}(x_c)\,\cA^{\mu\nu}$, due to the dressed
metric. That $\cA^*_{\mu\nu}=\cA_{\mu\nu}$ is rather  expected  in the
absence of boundaries, the photon action is invariant under (non-local)
dressing of the metric. 

If  two photons are entangled it means they  share a common origin at
$x_c$, where a non-factorisable polarisation entanglement matrix is acting e.g.
$\cE_{\rho\sigma}(x_c)\!=\! 1/\sqrt{2}\,(\epsilon_\rho^{(1)}\epsilon_\sigma^{(2)}\!-
\epsilon_\sigma^{(1)}\epsilon_\rho^{(2)})$. 
The correlation is then
\bea
C^*_{\alpha\beta}(x_1,x_2)=\int d^4x_c \sqrt{g^*}\,
\cA^*_{\alpha\rho} (x_1, x_c) \,g^{*\,\rho\sigma} \cE_{\sigma\lambda}(x_c)
\,g^{*\,\lambda\mu} \cA_{\mu\beta}^*(x_c; x_2)
\eea
%
The dressing factor, eq.(\ref{V}), in the metrics cancels out,
so in the absence of boundary conditions (i.e. detectors!), then
$C_{\alpha\beta}^*=C_{\alpha\beta}$,
and $C^{*\,\alpha\beta}(x_1,x_2)=V^{-2}(x_1) V^{-2}(x_2)\, C^{\alpha\beta}(x_1,x_2)$.

Assume now that  photon A interacts with a broken-phase environment of
the macroscopic detector  (defined by fixed lengths and scales) at point $x_k$, 
which {\it is} itself a local Riemannian geometry boundary condition
(b.c.)\footnote{rather than a ``mystical'' observer that collapses a wavefunction.}.
The photon action is then modified by the two boundaries at $x_k$ ($k=1,2$)
where detectors are placed
\bea\label{break}
\Delta S_k=\frac{1}{\langle\phi(x_k)\rangle^{4}}
\int d^4x \sqrt{g^*}\,  g^{*\,\mu\nu} J_\mu(x)\, A_\nu(x)\,
\frac{\delta^{(4)}(x-x_k)}{\sqrt{g^{*}(x)}}=
\frac{g^{* \,\mu\nu}(x_k)}{\langle\phi(x_k)\rangle^{4}}\, J_\mu(x_k) A_\nu(x_k)
 \eea

\medskip\noindent
where $J_\mu$ is a boundary current that interacts with the photon
  e.g. {\it local} fermions current $J_\mu=\overline\psi \gamma^a e_\mu^a \psi$
of  Weyl charge $-3/2-3/2+1=-2$.
Therefore $\Delta S_k$ has charge $-2$:
unlike in the absence of boundaries when photons correlator
was Weyl invariant, boundaries actions are not! Weyl
gauge symmetry is broken explicitly by non-vanishing Weyl charge of
$\Delta S_k$ and this is  due to localisation of the interaction
\footnote{Arguably, one could use the boundary metric
$g_{\mu\nu}$ in (\ref{break}) instead of $g^*_{\mu\nu}$, but the 
charge of $\Delta S_k$ remains non-zero (-4)
and the breaking still takes place, due to  localisation factor $\delta^{(4)}/\sqrt{g}$.}.
The local suppression scale  $\langle\phi(x_k)\rangle^4$ ensures a
dimensionless action. $\langle\phi(x_k)\rangle$ are field values in the
gauge used at each detector and to compare them one must transport say $\phi(x_2)$
to $x_1$: the result of doing so is denoted
$\phi(x_{2\ra 1})$, so the scales  mismatch ($r_s$) after doing so is
  \bea\label{r}
r_s=\frac{\phi(x_1)}{\phi(x_{2\ra 1})}= \frac{\phi(x_1)}{\phi(x_2) V(x_1)/V(x_2)}
=\frac{\phi^*(x_1)}{\phi^*(x_2)}=\exp\Big\{\int_{x_1}^{x_2} dy^\mu\, (\omega_\mu-\partial_\mu \ln\phi)\Big\}.
\eea

\medskip\noindent
This is Weyl gauge invariant and we notice the Weyl-Proca field $\omega_\mu^p\equiv
  \omega_\mu-\partial_\mu\ln \phi$  \cite{Ghilen0}
  under the integral\footnote{In
  other words the bilinear $\phi(x_2)^{-1} V(x_2,x_1) \phi(x_1)$ is gauge invariant with
 $V\!=\!\exp\int_{x_1}^{x_2} \omega$ (Wilson line of $\omega$)}.
The exponent of the relative factor $r_s$ is fixed by the net Weyl charge
of $\Delta S_k$ which is $-2$ giving a factor $r_s^{-2}$.
Thus, including the effect of the boundaries, we now have\footnote{
Additional effects can exist due to topological terms like
Euler term $G^*$ in action (\ref{e2}), which in the presence of boundaries
become relevant and bring extra couplings  not included here.}
\bea\label{cmunu}
C^*_{\mu\nu}(x_1,x_2)\propto V(x_1,x_2) C_{\mu\nu}(x_1,x_2),
\qquad V(x_1,x_2)=
\exp\Big\{2 \int_{x_2}^{x_1} \omega^p_\mu\,\, dz^\mu\Big\},
\eea

\medskip\noindent
so the Wilson line $V(x_1,x_2)=r_s^{-2}$ of Weyl-Proca field
is the correlation between the detectors.
Note the Wilson line connecting $x_1$-$x_2$ and
the geodesics paths of the two photons: $x_c$-$x_1$ and $x_c$-$x_2$,
enclose a triangle area; we expect that flux through this area
$\Phi=\int F_{\mu\nu} d\sigma^{\mu\nu}$ to bring an additional 
suppression factor $\exp(-\Phi)$ to
correlation $C_{\mu\nu}^*(x_1,x_2)$. Therefore in  eq.(\ref{cmunu}) 
we assumed that $x_c$ lies on the geodesic (\ref{Geo})
between the two detectors (i.e. a vanishing area)\footnote{
Minimising this area is also familiar in experiments \cite{QE1}.}.

Coincidence counting at the two detectors relies on the clocks
at the two detectors whose frequencies scale with their local mass scales.
In Weyl geometry the ratio of clocks frequencies is  the
ratio of their  scales  (\ref{r}) i.e. the Wilson line between
the detectors, so quantum fluctuations of $\omega_\mu$ make the relative
timing of the two detectors fluctuate. This provides the channel through
which the dressed geometry modulates the measured correlation.

Consider next the two measured polarisations vectors $v_\mu(x_1)$  and $u_\nu(x_2)$ at the
boundaries,  of the photon field $A_\mu$; to find their angle one must first parallel
transport, say, $u_\mu(x_2)$ to $x_1$ giving $\tilde u_\mu(x_{2\ra1})$, along a geodesic in this
geometry of $g_{\mu\nu}^*$; then one can compute their angle, 
then take the quantum average over quantum fluctuations due to $\omega$.
Since $g^*_{\mu\nu}$ is metric, it preserves norm and inner products under
parallel transport, just like in ordinary (local) Riemannian geometry of $g_{\mu\nu}$.
Since $g_{\mu\nu}$ and $g_{\mu\nu}^*$ are conformally related, the parallel transport
along their geodesics is then independent of $\omega_\mu$ and in the
definition of the angle ($\cos\theta$)
between the vectors the dependence on the dressing factor cancels out;
Weyl vector does not change the angle of polarisation vectors\footnote{Even if it did,
the coupling $\alpha\ll 1$ (in $\omega_\mu$) used here
would suppress $\omega_\mu$ dependence. Recall that $\overline\omega_\mu=0$.}.
As a result,
taking the quantum average (over $\omega_\mu$) of the total correlation
between the boundaries, the  dependence on the angle factors out
of the integral of this average; one finds for the final correlation
$\langle C^*\rangle^2_\omega=\cos^2\theta\langle V(x_1,x_2)\rangle_\omega^2$ where we denoted
\bea\label{QA}
\langle V(x_1,x_2)\rangle_\omega
\equiv \int \cD\omega_\mu^p\,\, V(x_1,x_2)\, \,\, e^{i S[\omega]}=
\exp\Big\{2\int_{x_2}^{x_1} dy^\mu \int_{x_2}^{x_1} dz^\nu
D_{\mu\nu}(y-z)\Big\},
\eea

\medskip\noindent
where  $\langle V(x_1,x_2)\rangle_\omega$ is the quantum average of  Wilson
line operator.
This result is the quantum effect of the Wilson line  in terms of the
propagator $D_{\mu\nu}$  of Weyl-Proca field; in the last step we used the
Proca action $S[\omega]$ for the Weyl-Proca field, working in the
unitary gauge i.e.  $\phi$=constant, $\omega_\mu^p=\omega_\mu$.
The correlation result in eq.(\ref{QA}) gives the same angle of the polarisations,
modulated by the Wilson line connecting the two boundaries,
$\langle V(x_1,x_2)\rangle_\omega$.

Let us discuss our results.
The measurement  of polarisation(s) {\it is} a boundary condition(s) (b.c.)
due to which the global mode
solution $C_{\mu\nu}^*(x_1,x_2)$ must satisfy additional constraints (b.c.)
since the detector at each boundary restricted the configuration space of
{\it global} operator \, $\Box^* \equiv g^{*\,\mu\nu}\nabla_\mu^* \nabla_\nu^*$
behind propagators $\cA^*_{\mu\nu}$.
The ``update'' at $x_2$ following a measurement at $x_1$ is not a causal signal - it is an
{\it algebraic} necessity to satisfy the global boundary condition(s)
of the dressed propagator(s).
This is a global-solution update, across a non-local (quantum) dressed manifold
($g_{\mu\nu}^*$), of its profile  at point $x_2$.
The ``non-locality'' of second photon (at $x_2$)  is simply the result of the
(b.c.) constraints by the detectors, via non-local $g^*_{\mu\nu}$ \footnote{
As a side-remark, in this picture, the wave
nature is the non-local spread  of the (non-local) field caused by the Wilson line
history (integral), while the particle point-like nature is at the specific
point (local Riemannian b.c.) where we chose to register the field interaction.
The particle is the local excitation while entanglement is 
geometric ``scaffolding'' provided by  gauge invariant non-local
$g_{\mu\nu}^*$ and the Wilson line operator present in it. }.

In other words what is called ``spooky action at distance''  is here just 
a feature of imposing  boundary conditions on a
global  differential (dressed) operator $\Box^*$ in (\ref{dme})  - an
operator of structure  imposed by {\it  Weyl
  gauge invariance} which requires the use of $g^*_{\mu\nu}$; it is not an action
{\it through} flat spacetime as interpreted in quantum mechanics,
but an action {\it intrinsic} to the structure of the {\it quantum non-local}
spacetime ($g_{\mu\nu}^*$) - recall that Wilson line (in $g^*_{\mu\nu}$) is
a sum of  quantum fluctuations
of $\omega_\mu$ with background value $\overline\omega_\mu=0$ as demanded by Lorentz invariance.
The two states do not exchange information  faster-than-speed-of-light;
instead, this information is encoded in the  {\it gauge invariant}
(global, quantum-dressed)  metric.
A physical observer can only measure Weyl gauge invariant (dressed) variables,
so our explanation of ``spooky action''  is  enforced by this gauge symmetry
as  seen through our-world Riemannian geometry of $g_{\mu\nu}^*$.
 Entanglement is then a ``shadow'' of a shared gauge-history
 (at $x_c$)  in gauge invariant geometry ($g_{\mu\nu}^*$).
If we accept Weyl gauge symmetry, we live in a
 gauge invariant (dressed) non-local geometry and entanglement
 is the ``tax'' paid for that and evidence of this underlying  symmetry.

 What happens when $\omega_\mu$ acquires mass (see section~\ref{2})? The Wilson
 line quantum average $\langle V(x_1,x_2)\rangle_\omega$ will be suppressed
 and decoherence takes place, see (\ref{QA}).
 For a distance of order $d\sim1200$ km over which entanglement was tested \cite{QE1,QE2}
 a mass $m_\omega=2\pi\hbar/\lambda\sim 1.64\, (2\pi) \times  10^{-13}$ eV is found
 for $\lambda\sim d$, before decoherence takes place due to suppression by
 $m_\omega$\footnote{This bound is found in static case, giving 
 with $D_{\mu\nu}\propto \exp(- m_\omega \vert x -y\vert)/(4\pi\vert x-y\vert)$ from (\ref{QA}).}.
 Such  value of $m_\omega$ is  possible (section~\ref{2}) for
 Weyl gauge coupling of order $m_\omega/M_P\!\sim\! 10^{-40}$ which
 avoids current experimental constraints.
 A vector boson of mass close to $10^{-13}$ eV was also found
 \cite{ER} for a dark matter
 candidate which in Weyl geometry is actually $\omega_\mu$!
 Bounds on $m_\omega$ come from atomic clocks
 deviations as probes of non-metricity, with a value again near  $10^{-13}$ eV
 \cite{saridakis} and from  black-hole superradiance \cite{caputo,oscar}.
 It is   interesting that these results seem to correlate, in this mass region,
 entanglement by $\omega_\mu$, atomic clock tests of non-metricity due to
 $\omega_\mu$, dark matter of which $\omega_\mu$ is the candidate in WQG,
 and black-hole superradiance.
 
 This explanation, based on Weyl gauge symmetry, may  be a step towards reconciling
 foundational quantum conflicts, by a QFT {\it re-interpretation of
 state vectors by a (deterministic) global boundary-value problem in Weyl gauge invariant
 quantum-non-local geometry.} This seems to: 1) resolve
 Einstein's non-locality concern by replacing ``spooky action''
by global algebraic constraints required by Weyl gauge invariance;
2) it bypasses Bell's theorem \cite{Bell} by {\it non-local} hidden variables in the metric,
with non-locality as a  purely  {\it quantum} effect (sum of 
quantum fluctuations of $\omega_\mu$) and 3): it formalises Bohr's
concept of experimental indivisibility by treating the measurement apparatus
as a global boundary condition, which dynamically shapes the non-local geometry
of the dressed metric.
This  approach to entanglement  is conceptually different from
ER=EPR \cite{EPR} of local Riemannian space; unlike the latter which
uses string theory and AdS/CFT, the present case
is considerably simpler,  based solely on Weyl gauge invariance in 4D geometry,
and demands that observables respect this symmetry, which enforces
quantum non-locality and non-commutativity\footnote{
in the vector space of these gauge invariant operators/observables.}.

Finally, recall that behind this, at the  most fundamental level, there is
actually an underlying action that is completely  local and Weyl gauge
invariant, defined by Weyl geometry, and with no lengths and  masses.
Quantum non-locality  emerges  when we define physical i.e. (Weyl) gauge-invariant
observables, in the vector space of dressed (non-local, gauge invariant) metric and
fields of our Riemannian world. Space-time metric $g^*_{\mu\nu}$ is an
emergent property,  of connected nodes by Wilson lines, and entanglement seems
part of it, if our study of entanglement is correct;
if so, roughly speaking, entanglement ``is'' (part of) dressed,
Weyl gauge invariant 4D geometry!

\section{Conclusions}

In this work we studied  Weyl conformal geometry and its Weyl gauge symmetry
from a more familiar Riemannian view that is however manifestly
Weyl-gauge {\it invariant}.
Our interest in Weyl geometry is motivated by the
fact that its  action, the  Weyl quadratic (gauge theory of) gravity and
especially the more general Weyl-Dirac-Born-Infeld (WDBI)
action, are  good candidates for quantum gravity.
These are gauge theories of the Weyl group,  Weyl anomaly-free, with 
exact geometric interpretation and all mass scales of geometric origin.
In the Stueckelberg broken phase, when $\omega_\mu$ becomes massive and decouples,
Weyl geometry/connection become Riemannian and Einstein-Hilbert gravity is
recovered below Planck scale together with $\Lambda>0$.
Moreover, the  WDBI action of Weyl geometry is Weyl gauge invariant
in arbitrary $d$ dimensions;
this action is, to our knowledge, the only theory that  does not need a UV
regulator! In $d=4-2\epsilon$
it includes non-perturbative corrections  to the (geometrically) regularised
Weyl quadratic gravity which is  its leading order.
All this explains our interest in Weyl geometry.

For a better physical understanding of Weyl gauge symmetry that defines Weyl
geometry, it  helps to have a more familiar Riemannian geometry view
in which we actually live. Normally, the Riemannian view is used to
describe the  low-energy (effective theory of the)
broken phase, in which Weyl connection becomes Riemannian and
Einstein-Hilbert action is recovered. This view is in a sense
a projective representation of the underlying Weyl geometry, its gauge symmetry
and action. Regarding  the symmetric phase,
ordinary Riemannian curvatures  transform in a non-covariant, complicated way.
The question is whether we can also have a Riemannian picture 
that is manifestly {\it Weyl gauge invariant} i.e. physical.  That would be more like an
isomorphic representation of the Weyl geometry picture, ``translated'' in
Riemannian geometry ``language''.
We showed this is indeed possible using  Weyl gauge invariant 
{\it non-local} dressed fields (metric, curvatures) with  dressing due to 
Wilson line  of dilatations.  The  dressed fields have the advantage
that  their use gives automatically gauge invariant i.e.
physical results (that a Riemannian physical observer
is measuring, like masses, etc); their use is in fact {\it necessary} here, to avoid
gauge-dependent artefacts (in e.g. absence/presence of metric singularities, etc).

We found that Weyl geometry (connection) can be mapped into Riemannian geometry
(connection) of the dressed {\it non-local} metric, at the cost of (gauge-induced)
non-commutativity in the ultraviolet, due to the (derivative of) Wilson line
\footnote{More exactly, 
this map is trading a Weyl geometry affine non-metric formulation
($g_{\mu\nu}$, $\tilde\nabla_\mu(\tilde\Gamma)$)
or its non-affine
gauge-covariant metric formulation ($g_{\mu\nu}$, $\hat\nabla_\mu(\tilde\Gamma)$)
for a non-local, gauge-invariant, quantum-dressed metric geometry
with non-commuting derivatives on dressed fields, on Riemannian side.}.
This is an important result.
This non-commutativity proportional to Weyl flux ($\propto F_{\mu\nu}$)
is a quantum effect since background $\overline\omega_\mu=0$, and is present
above the scale $m_\omega$ at which Weyl gauge symmetry is broken. This breaking
generates $M_p$ and $m_\omega$ which are gauge invariant in the dressed picture,
being set by $\langle\phi^*\rangle$. In this sense, $1/M_p$ is the shortest
{\it physical} length scale of the theory, playing the role of a physical
cutoff of its low-energy (Einstein-Hilbert) description.

  Dressed fields are non-local, and so are their interactions, which are spread
along Wilson-line paths instead of being point-like. In (perturbative) Weyl quadratic gravity
this may soften the UV behaviour
in a way reminiscent of non-local string interactions, but without strings use.
The WDBI action follows a different, non-perturbative logic
and is closer in spirit to D-branes effective
actions, with perturbative Weyl quadratic gravity as
its leading order in $\xi^2$. As in strings case, a resolution bound that may
exist in perturbative regime need not hold non-perturbatively, so for WDBI the scale
$1/M_p$ can be seen as the scale where perturbative description
changes, not as a bound on resolution, since the WDBI action needs no UV regulator!
Finally, the Born-Infeld type condition for  $\det>0$ in WDBI action restricts $F_{\mu\nu}$
relative to curvature terms, which may oppose gravitational collapse
and the formation of point-like singularities in the
dressed metric $g_{\mu\nu}^*$
and possibly lead to a bounce. This is a conjecture that must be tested
on an explicit solution for $g_{\mu\nu}^*$.

 The Wilson line is driven
 by quantum fluctuations of $\omega_\mu$ ($\overline \omega_\mu=0$);
 so non-local dressing and non-commutativity are quantum effects, as mentioned.
 While Weyl geometry picture is fundamental and local,  non-locality 
 in  Riemannian picture of dressed fields appears  as a  representation
 artefact    of ``translating'' (local)  Weyl geometry and Weyl gauge covariant fields
 into our  Riemannian geometry  of Weyl {\it gauge invariant} fields and
 observables like masses  that a (Riemannian) observer is measuring.
 Such effective non-locality is thus {\it physical},
 woven in our (dressed) geometry of gauge invariant
 $g_{\mu\nu}^*$, and  is indirect evidence of Weyl gauge symmetry!

 The aforementioned non-local map from Weyl to Riemannian  geometry
 extends to their associated actions. We  found a non-local map
 from Weyl quadratic action in $d$ dimensions to
a quadratic  action of exactly same form in Riemannian geometry of
dressed $g_{\mu\nu}^*$ but with  UV non-commutativity.
Both actions are Weyl gauge invariant in $d$ dimensions!
The map also works for the more fundamental  WDBI
action which is Weyl gauge invariant in  arbitrary $d$ dimensions in both geometries.
Unlike for  the metric, the equation of motion of  $\omega_\mu$ 
does not commute with the dressing of the metric,
which underlines the fundamental difference between Riemann and Weyl geometries,
that we explained.
At lower energies Weyl gauge symmetry is broken,  massive $\omega_\mu$
decouples ($F_{\mu\nu}\!=\!0$),  and point-like physics returns, dressing
becomes irrelevant and commutativity is  restored; the  Riemannian picture of
 (high energy)  dressed fields recovers  that of
 low-energy broken phase of ordinary fields and of Einstein-Hilbert gravity.

 As an application, we argued that it is possible that
 the quantum non-local connection in
 the entanglement of two photons in quantum mechanics, be explained conceptually
 in quantum field theory in curved space-time, by the
 non-local  (quantum-dressed) metric  $g^*_{\mu\nu}$.
 If so, this connection would be a
 footprint   of Weyl gauge  symmetry, that appears into  our-world of
 non-local  quantum-dressed Riemannian geometry.
 This may be possible only for ultralight Weyl gauge bosons
 (ultraweak $\alpha\!\ll\! 1$), as opposed to the more conservative view
 where the mass of $\omega_\mu$ is assumed large (both cases evade current  experimental
 constraints).

 To conclude, these results show a {\it non-local} map and dual description
 of Weyl versus Riemannian geometries and of their actions in the symmetric phase,
 leading to better understanding of Weyl gauge symmetry and of its footprints
 in a  more familiar Riemannian perspective that is manifestly
 Weyl gauge invariant!

 \vspace{1cm}
 \begin{center}
   ------------------------------------
   \end{center}

\newpage
\section{Appendix}

\def\theequation{A-\arabic{equation}}
\def\thesubsection{A}
\setcounter{equation}{0}
\def\thefigure{A-\arabic{figure}}
\def\thelabel{A}

\subsection{Weyl geometry formulae}\label{A}

We present some  formulae in Weyl geometry and the
relation to Riemannian geometry, in arbitrary $d$ dimensions.
The  curvature tensors 
in various formulations of Weyl geometry and the relations
to their counterparts in Riemannian geometry are found
in \cite{DG1} (Appendix) and \cite{AC,CDA2}. From eq.(\ref{curvature-1})
\medskip
\begin{equation}
\tilde R^\rho{}_\sigma{}_{\mu \nu} = \partial_\mu \tilde \Gamma^\rho_{\nu \sigma}
  - \partial_\nu \tilde \Gamma^\rho_{\mu \sigma} + \tilde \Gamma^\rho_{\mu \tau}
  \tilde \Gamma^\tau_{\nu \sigma} - \tilde \Gamma^\rho_{\nu \tau} \tilde \Gamma^\tau_{\mu \sigma},
\label{curvature-A}
\end{equation}
and with notation
\medskip
\bea
\tilde R_{\alpha\mu\nu\sigma}= g_{\rho\alpha}\tilde R^\rho_{\,\,\,\mu\nu\sigma},
\qquad
\tilde R_{\mu\sigma}=\tilde R^\rho_{\,\,\,\mu\rho\sigma},
\qquad
\tilde R= g^{\mu\sigma} \,R_{\mu\sigma},
\eea
one obtains  (using (\ref{tGamma}))
\bea\label{ss1}
\tilde R_{\alpha\mu\nu\sigma}
\!\!\!&=&
\!\!\!
R_{\alpha\mu\nu\sigma}
+\Big\{g_{\alpha\sigma} \nabla_\nu\w_\mu- g_{\alpha\nu} \nabla_\sigma \w_\mu
-g_{\mu\sigma}\nabla_\nu \w_\alpha +g_{\mu\nu} \nabla_\sigma\w_\alpha
+g_{\alpha\mu}\, F_{\nu\sigma}
\Big\}
\nonumber\\
&+&\Big\{ \w^2 (g_{\alpha\sigma} g_{\mu\nu}-g_{\alpha\nu} g_{\mu\sigma} )
+\w_\alpha \,(\w_\nu g_{\sigma\mu}-\w_\sigma g_{\mu\nu})
+\w_\mu (\w_\sigma g_{\alpha\nu} -\w_\nu g_{\alpha\sigma})\Big\},\quad
\nonumber\\
\tilde R_{\mu\sigma}\!&=&\!\!\! R_{\mu\sigma} 
\!+ \Big[\frac12\, d\, F_{\mu\sigma}-(d-2)\nabla_{(\mu} \omega_{\sigma)}
 - g_{\mu\sigma} \nabla_\lambda\omega^\lambda\Big]
\!+(d-2) (\omega_\mu\omega_\sigma
-g_{\mu\sigma} \omega_\lambda\omega^\lambda),
\nonumber\\[7pt]
\tilde R &=& 
R-2 (d-1)\, 
\nabla_\mu \omega^\mu -(d-1) (d-2) \,\omega_\mu \omega^\mu.
\eea
The rhs of (\ref{ss1}) is in Riemannian notation, with 
 $\nabla_\mu\omega_\nu=\partial_\mu \omega_\nu -\Gamma_{\mu\nu}^\rho\omega_\rho$, and
$\Gamma$ the Levi-Civita connection:
$\Gamma_{\mu\nu}^\rho(g)=(1/2)\,g^{\rho\lambda} (\partial_\mu g_{\nu\lambda}+\partial_\nu g_{\mu\lambda}
-\partial_\lambda g_{\mu\nu}).$

The relation to ``hat'' formulation of eq.(\ref{curvature-hat}) is
shown in eq.(\ref{relations-curvatures})
\bea\label{th}
\hat R_{\alpha\mu\nu\sigma} =\tilde R_{\alpha\mu\nu\sigma} -g_{\alpha\mu} \,F_{\nu\sigma},
\qquad
\hat R_{\mu\sigma}=\tilde R_{\mu\sigma}-F_{\mu\sigma},
\qquad
\hat R=\tilde R,
\eea

\medskip\noindent
with $\hat R_{\alpha\mu\nu\sigma}\!=\! g_{\alpha\lambda} \hat R^{\lambda}_{\,\,\,\,\mu\nu\sigma}$.
Here $R_{\alpha\mu\nu\sigma}=g_{\alpha\lambda}  R^\lambda_{\,\,\,\mu\nu\sigma}$,
$ R_{\mu\nu}= R^\lambda_{\,\,\,\mu\lambda\nu}$, $ R=g^{\mu\nu} \,R_{\mu\nu}$ are the Riemann and
Ricci tensor and scalar  of  Riemannian geometry defined by $\Gamma(g)$,
in $d$ dimensions.

We have  $\hat R_{\mu\nu}-\hat R_{\nu\mu}= 
(d-2) \hat F_{\mu\nu}$, so $\hat R_{\mu\nu}$ is not symmetric if $d\not=2$.
The field strength $ F_{\mu\nu}=\partial_\mu\omega_\nu-\partial_\nu\omega_\mu=\hat F_{\mu\nu}$
 has the same expression in Weyl  and Riemann geometries.

In the Weyl gauge covariant formulation 
the Weyl tensor $\hat C^\mu_{\,\,\nu\rho\sigma}$ associated to the  Riemann tensor
$\hat R^\mu_{\,\,\,\,\nu\rho\sigma}$ is  equal to its Riemannian counterpart
($C^\mu_{\,\,\nu\rho\sigma}$) \cite{DG1} (eq.A-25)
\bea\label{ccc}
\hat C^\mu_{\,\,\nu\rho\sigma}=C^\mu_{\,\,\nu\rho\sigma}.
\eea

\medskip
The following identities of Weyl conformal geometry (in the ''hat'' notation) 
exist and are similar to  those of Riemannian geometry, but in a Weyl gauge covariant
form \cite{DG1}, \cite{CDA2}
\medskip
\bea\label{hatG}
\hat G=\hat R_{\mu\nu\rho\sigma}\,\hat R^{\rho\sigma\mu\nu} - 4 \,\hat R_{\mu\nu}\hat R^{\nu\mu}
+\hat R^2,
\eea
and
\bea
\hat C_{\mu\nu\rho\sigma}^2=\hat R_{\mu\nu\rho\sigma}\,\hat R^{\rho\sigma\mu\nu}-\frac{4}{d-2}\,
\hat R_{\mu\nu}\,\hat R^{\nu\mu}
+\frac{2}{(d-1)(d-2)}\,\hat R^2.
\eea
This gives
\bea\label{hatRmunu}
- \hat R_{\mu\nu}\,\hat R^{\nu\mu}+ \frac{d-2}{4\,(d-3)} \,(\hat C_{\mu\nu\rho\sigma}^2-\hat G)
+\frac{d}{4\,(d-1)}\,\hat R^2=0.
\eea

\medskip
This equation can be used to express any of these terms in terms of the rest.
This relation is also valid in Riemannian geometry (without a hat), obvious by setting
$\omega_\mu=0$:
 \bea\label{Rmunu}
  - R_{\mu\nu}\, R^{\nu\mu}+ \frac{d-2}{4\,(d-3)} \,(C_{\mu\nu\rho\sigma}^2- G)
 +\frac{d}{4\,(d-1)}\, R^2=0.
 \eea
Last two equations together with (\ref{rel}) to (\ref{rl}) and  (\ref{ccc})
are used to find the relation in eq.(\ref{GG}).
We also have
\bea
\hat R^\rho{}_\sigma{}_{\mu \nu}=\tilde R^\rho{}_\sigma{}_{\mu \nu}-\delta_\sigma^\rho \,F_{\mu\nu}
=
R^\rho{}_\sigma{}_{\mu \nu}(g^*)-\delta_\sigma^\rho \,F_{\mu\nu}.
\eea
With (\ref{ss1}) then
\begin{align}
  &  \hat R_{(\mu\sigma)}=\tilde R_{(\mu\sigma)}\!=
R_{(\mu\sigma)}(g^*)= R_{\mu\sigma}(g)-
g_{\mu\sigma} \nabla_\lambda\omega^\lambda 
\!+(d-2) (\omega_\mu\omega_\sigma
-g_{\mu\sigma} \omega_\lambda\omega^\lambda -\nabla_{(\mu} \omega_{\sigma)}),
\nonumber
\\[7pt]
& \hat R=\tilde R=R(g^*)=R(g)-2 (d-1)  \nabla_\lambda\omega^\lambda
- (d-1) (d-2) \,\omega_\lambda\,\omega^\lambda.
\end{align}

\bigskip

\def\theequation{B-\arabic{equation}}
\def\thesubsection{B}
\setcounter{equation}{0}
\def\thefigure{B-\arabic{figure}}
\def\thelabel{B}

\subsection{{Derivative of the Wilson line}}\label{B}

Consider a path $\gamma: y^\mu(\lambda)$ as defined after eq.(\ref{auto})
with endpoint $x^\mu=y^\mu (1)$ ($y^\mu(0)=-\infty$).
We  compute the  variation of the line integral below
\bea
\cJ(\gamma, x)=\int_{-\infty}^{x} \w_\mu \,dy^\mu
\equiv\int_0^1 \w_\mu \,\frac{dy^\mu}{d\lambda} \,d\lambda,\qquad\quad V=\exp \cJ.
\eea

\medskip\noindent
The total variation   is (using $\delta d=d \delta$)
\medskip
\bea
\delta \cJ(\gamma,x)&=&
\int_0^1 \,\delta\Big( \w_\mu \,\frac{dy^\mu}{d\lambda}\Big) d\lambda
=
\int_0^1 \Big\{\partial_\nu \w_\mu \delta y^\nu \frac{d y^\mu}{d\lambda}+
\w_\mu \frac{d\delta y^\mu}{d\lambda}\Big\} d\lambda
\\
&=&
\w_\mu \delta y^\mu \Big\vert_0^1
+\int_0^1\partial_\nu \w_\mu \frac{dy^\mu}{d\lambda} \delta y^\nu d\lambda
-\int_0^1\frac{d\w_\mu}{d\lambda} \delta y^\mu d\lambda
\\
&=&
\w_\mu \delta y^\mu \Big\vert_0^1
+\int_0^1\partial_\mu \w_\nu \frac{dy^\nu}{d\lambda} \delta y^\mu d\lambda
-\int_0^1 \partial_\nu \w_\mu\frac{dy^\nu}{d\lambda} \delta y^\mu d\lambda
\\
&=&
\w_\mu \delta y^\mu \Big\vert_0^1
+\int_0^1 F_{\mu\nu} \frac{dy^\nu}{d\lambda} \delta y^\mu\, d\lambda.
\eea
Therefore
\bea\label{global}
\delta \cJ(\gamma,x)=
\delta \int_{y(\lambda)} \w_\mu dy^\mu=
\w_\mu \,\delta x^\mu  +\int F_{\mu\nu} \, dy^\nu\, \delta y^\mu,
\qquad (\, d y^\nu=y^{\nu\,\prime}(\lambda) d\lambda\,).
\eea
%
This is  the total variation (differentiation) of $\cJ(\gamma; x^\mu)$:
first term is due to  local variation around endpoint $x^\mu$
for $\gamma$ fixed;
 second term  is due to global variation (deformation) of path $\gamma$ and
measures the flux through the narrow strip area swept by  deformation $\delta y^\mu$,
bordered by the path  $y^\nu(\lambda)$ and  $\delta y^\mu$.
The  local variation (which takes place ``at the tip of the path'',
$\gamma$ fixed, $\delta y^\mu=0$), gives the partial derivative with
respect to  endpoint $x$ 
\bea\label{ac}
\frac{\partial \cJ}{\partial x^\alpha} = \w_\alpha, \qquad x^\alpha\equiv y^\alpha(1).
\eea

\medskip
 From (\ref{ac}) then $\partial_\mu V=\omega_\mu V$ and eq.(\ref{com})  follows.
 This  result for  Wilson lines is known 
 since pioneering works of \cite{terning,mandelstam,mandelstam2}
but  some  clarifications are necessary.
  In (\ref{ac}) $\partial_\mu\equiv\partial/\partial x^\mu$ is  not a
new differential operator, it is the ordinary partial
derivative in space-time. The autoparallel  curve $y(\lambda)$ (\ref{auto}) fixes
only the derivative along its tangent
$d \cJ/d\lambda=\omega_\mu  (dy^\mu/d\lambda)$.
But in general, one must clarify if there is a  local or
global path deformation (giving $\partial_\mu \cJ$ and total differential
$d\cJ$, respectively).

A local deformation giving
 eq.(\ref{ac}) (``tip rule'') means that the nearby path is the old path
($\delta y^\mu=0$) plus a short segment, along $y^\mu(\lambda)$.
This is the minimal (local) choice available:
only the tip rule makes $\partial_\mu \cJ$ a local quantity:
by definition, $\partial_\mu$ is a local operator, so it  depends only
on data in the neighbourhood $\cO(\epsilon)$ around
the endpoint $x$. Deformations $\delta y^\mu$
near this point in all directions, parallel or transverse to the path, must 
be small $\cO(\epsilon)$. While the first term in (\ref{global}) is
$\cO(\epsilon)$, the second term is $\cO(\epsilon^2)$  {\it
  in the neighbourhood of $x$}
and is suppressed by $\epsilon$ relative to first term
when computing   $\partial\cJ/\partial x^\mu$
by  the limit  $\epsilon\!\ra \!0$ (as in $f'(x)\!=\![f(x+\epsilon)\!-\!f(x)]/\epsilon$);
its contribution vanishes in this limit, and (\ref{ac}) follows\footnote{This  is also
seen in a discrete case $\cJ\!\approx \!\sum_{i=1}^N \omega_\mu(y_i) (y_i-y_{i-1})^\mu$.
For endpoint  $y_N^\mu\equiv x^\mu$, $\partial \cJ/\partial y_N^\mu=\omega_\mu+\cO(\epsilon)$,
while for interior points $j\!\not=\!N$, 
$\partial \cJ/\partial y_{j}^\mu=F_{\mu\nu}(y_{j})\, \epsilon^\nu=
\cO(\epsilon)$, with  $(y_i-y_{i-1})^\mu=\epsilon^\mu$
hence (\ref{ac}).}. This is simply the definition of partial derivative
$\partial_\mu \cJ(\gamma;x)$: a variation of $\cJ$ with respect to
$x$ with  $\gamma$ fixed\footnote{
This  is consistent with a second, physical  argument (causality) that
  the history up to $x^\mu$ is not re-written when  the endpoint $x$ of $y(\lambda)$
  moves under $\partial_\mu$, as detailed in section~{3.1}.}$^,$\footnote{
It is exactly the locality of $\partial_\mu$  (the tip rule  for path deformation)
of dressed fields that enables one
to map Weyl to Riemannian geometry and their Lagrangians and equations (section~\ref{3.4}).
This  is because dressing  commutes with the covariant differentiation of the fields:
assume that $\partial_\mu \ln V=\omega_\mu + a_\mu$, with $a_\mu$ due to
path history (non-local) deformation and flux integral in (\ref{global}).
Then for a scalar $T$ of charge $\tilde q$:
$\hat\nabla_\mu T^*=\partial_\mu T^*= V^{\tilde q}
(\hat\nabla_\mu T+ \tilde q \, a_\mu T)$, so
$\hat \nabla_\mu T^*=(\hat \nabla_\mu T)^*+ \tilde q a_\mu T^*$.
Therefore, $\hat \nabla_\mu T^*=(\hat \nabla_\mu T)^*
=\nabla_\mu^* T^*$ and dressing commutes with covariant  differentiation
if and only if $a_\mu=0$ i.e. $\partial_\mu V=\omega_\mu V$
(tip rule).}.\!
The history of $V(x)$ is the autoparallel (\ref{auto}) and $\partial_\mu$
acts on $V(x)$ along this given history extending it at its tip. An autoparallel
ending at $x\!+\! dx$ is a slightly different curve and  the $F_{\mu\nu}$ flux through
the strip between the two is the extra term in  (\ref{global})
and in $d\cJ$ (\ref{B8}).

By contrast,  when the whole path is deformed  under $x\ra x+dx$, eq.(\ref{global}), we
have a total differential
\bea\label{B8}
\frac{d}{dx^\mu}\, J\big(y(x), x\big)=\partial_\mu \cJ
+\int_0^1\frac{\delta \cJ}{\delta y^\nu} \, \frac{\partial y^\nu}{\partial x^\mu}\, d\lambda.
\eea
 $\delta\cJ/\delta y^\nu$ is the interior\footnote{so that the contribution
of the endpoint $x^\mu\!=\!y^{\mu}(1)$ is not counted twice, as
this is already included in $\partial_\mu \cJ$.} functional derivative,
$\delta\cJ/\delta y^\nu\!=\!
F_{\nu\alpha} \, y^{\alpha \,\prime} (\lambda)$, and
second term is  the flux integral in (\ref{global})
(scale holonomy); (\ref{B8}) is (\ref{global}) evaluated on 
variation $\delta y^\mu=(\partial y^\mu/\partial x^\nu)\delta x^\nu$.

 The local deformation case  accounts for holonomy,
 via (local) non-commutativity (\ref{com}) on (non-local) dressed fields:
 each tip rule extension in (\ref{com}) extends  the history {\it locally}
 by $dx^\mu$ or  $dx^\nu$ in different directions,
giving two different paths to $x+dx^\mu+dx^\nu$ bounding the plaquette; then the two
partial derivatives ($\partial_\mu$, $\partial_\nu$)
in different directions do not commute anymore,
and  $F_{\mu\nu}$ is a local manifestation, around the endpoint $x$,
of scale holonomy, as a flux through the plaquette of area $dx^\mu dx^\nu$ around
endpoint $x$.
$V(x)$ is  a path functional only through its history 
(different histories give different $V(x)$ for $F_{\mu\nu}\!\not=\!0$). 
Relation (\ref{com}) is a statement that $V(x)$
(or a dressed field) is not a function of $x$ alone but depends on the path.

\def\theequation{C-\arabic{equation}}
\def\thesubsection{C}
\setcounter{equation}{0}
\def\thefigure{C-\arabic{figure}}
\def\thelabel{C}

\subsection{Wilson line path integral}\label{AC}

For action (\ref{fwg}) we give the exact mathematical definition of 
 Wilson line integral (\ref{wd}), with the path of integration $x^\mu(\lambda)$
in spherical coordinates $x^\mu\equiv (t,r,\theta,\phi)$ for
$(\omega_0, \omega_1,\omega_2=0,\omega_3=0)$ independent of $t$.
Then the dressed metric becomes
\medskip
\bea
g_{\mu\nu}^*(x)=g_{\mu\nu}(x)\,\exp\Big\{2\int_{-\infty}^x \big(
\omega_0\, dt+\omega_1 \, dr \big)\Big\}=
g_{\mu\nu}(x) \phi^2(x)\,\exp\Big\{2\int_{-\infty}^x 
\omega_0\, dt \Big\}.
\eea

\medskip\noindent
The equation of motion for  $\omega_1(r)$ gives
$\omega_1(r)=(1/2)\partial_1\ln\phi^2(r)$ while for $\omega_0(r)$:
\medskip
\bea
\partial_1^2 \omega_0+(\partial_1\omega_0)\,\,\partial_1
\ln \Big[ \frac{r^2}{(-g_{00}\,g_{11})^{1/2}}\Big]
-\frac{6\alpha_1}{\alpha_3}\,\omega_0\,\,\phi^2(r)\,\,g_{11}=0
  \eea

  \medskip\noindent
  The path of integration is given by the geodesic
  \bea
  X^\nu \hat\nabla_\nu X^\mu=0,\qquad {\rm{or}}\qquad
  X^\nu \,(\tilde\nabla_\nu -\omega_\nu)\,X^\mu =0,
  \qquad X^\mu\equiv\frac{\partial x^\mu}{\partial \lambda},\,\,\, x^\mu=e_a^\mu x^a.
  \eea
  which gives
  \bea
  \frac{\partial^2 x^\mu}{\partial\lambda^2}-\omega_\alpha\,\,
  \frac{\partial x^\alpha}{\partial \lambda}\frac{\partial x^\mu}{\partial\lambda}
  +\tilde\Gamma^\mu_{\alpha\beta} \,\,\frac{\partial x^{\alpha}}{\partial\lambda}\,
  \frac{\partial x^{\beta}}{\partial \lambda}=0.
  \eea

  \medskip\noindent
  Next, replacing $\tilde\Gamma$ in spherical coordinates,
set $\omega_{2,3}=0$, $\theta=\varphi=$constant for convenience
  and   $t=t(\lambda)$, $r=r(\lambda)$ we find from last equation
  \medskip
  \bea
t''(\lambda)+r'(\lambda)\, t'(\lambda)
+\Big[\omega_1(r)+\partial_1(\ln g_{00})\Big]-\frac{g_{11}}{g_{00}}
\,\omega_0\,  r^{\prime \,\,2}(\lambda)&=&0,\qquad\qquad
\\[4pt]
r''(\lambda)+r'(\lambda)\,t'(\lambda)\,\omega_0-t^{\prime\,2}(\lambda)\,\,
\frac{g_{00}}{g_{11}}\,\Big[\frac{1}{2}(\partial_1\ln g_{00})+\omega_1\Big]
  +\frac{1}{2}\,r^{\prime \,2}(\lambda)\,\partial_1(\ln (-g_{11}))&=&0.
  \eea
  where  $\omega_1(r)=\frac12\partial_1 \ln\phi^2(r)$.
  The Wilson line integral may be evaluated analytically
  under additional simplifying assumptions.

\end{document}